\documentclass[trackchanges,twocolumn]{aastex7}
\usepackage{multirow}
\usepackage{graphicx} 
\usepackage[table]{xcolor}
\usepackage{rotating}

\definecolor{vFifty}{HTML}{DDEBFF}
\definecolor{vTwoHundred}{HTML}{E5F4E3}
\definecolor{vFiveHundred}{HTML}{FFF1CC}
\definecolor{vThousand}{HTML}{F9E0E7}
\newcolumntype{F}{>{\columncolor{vFifty}}c}
\newcolumntype{H}{>{\columncolor{vTwoHundred}}c}
\newcolumntype{Q}{>{\columncolor{vFiveHundred}}c}
\newcolumntype{T}{>{\columncolor{vThousand}}c}

\begin{document}

\title{Resolving a Clumpy Disk Wind in 4U 1957+115 with XRISM}

\correspondingauthor{Paul A. Draghis}
\author[0000-0002-2218-2306]{Paul A. Draghis}
\altaffiliation{MIT Kavli Postdoctoral Fellow}
\affiliation{Department of Physics \& Kavli Institute for Astrophysics and Space Research, Massachusetts Institute of Technology, 70 Vassar St, Cambridge, MA 02139, USA}
\email[show]{pdraghis@mit.edu}

\author[0000-0003-0172-0854]{Erin Kara}
\affiliation{Department of Physics \& Kavli Institute for Astrophysics and Space Research, Massachusetts Institute of Technology, 70 Vassar St, Cambridge, MA 02139, USA}
\email{ekara@mit.edu}

\author[0000-0001-9735-4873]{Ehud Behar}
\affiliation{Department of Physics, Technion, Technion City, Haifa 3200003, Israel}
\email{behar@physics.technion.ac.il}

\author[0000-0003-2869-7682]{Jon M. Miller}
\affiliation{Department of Astronomy, University of Michigan, 1085 South University Avenue, Ann Arbor, MI 48109, USA}
\email{jonmm@umich.edu}

\author[0000-0002-7303-8144]{Vera Berger}
\affiliation{Department of Physics \& Kavli Institute for Astrophysics and Space Research, Massachusetts Institute of Technology, 70 Vassar St, Cambridge, MA 02139, USA}
\email{vlberger@mit.edu}  

\author[0000-0002-0568-6000]{Joheen Chakraborty}
\affiliation{Department of Physics \& Kavli Institute for Astrophysics and Space Research, Massachusetts Institute of Technology, 70 Vassar St, Cambridge, MA 02139, USA}
\email{joheen@mit.edu}  

\author[0000-0003-4511-8427]{Peter Kosec}
\affiliation{Center for Astrophysics $\vert$ Harvard \& Smithsonian, 60 Garden St., Cambridge, MA 02138, USA}
\email{peter.kosec@cfa.harvard.edu}  

\author[0009-0001-9034-6261]{Christos Panagiotou}
\affiliation{Department of Physics \& Kavli Institute for Astrophysics and Space Research, Massachusetts Institute of Technology, 70 Vassar St, Cambridge, MA 02139, USA}
\email{cpanag@mit.edu}  

\author[0000-0002-2381-4184]{Swati Ravi}
\affiliation{Department of Physics \& Kavli Institute for Astrophysics and Space Research, Massachusetts Institute of Technology, 70 Vassar St, Cambridge, MA 02139, USA}
\email{swatir@mit.edu}  

\author[0000-0003-0931-0868]{Fatima Zaidouni}
\affiliation{Department of Physics \& Kavli Institute for Astrophysics and Space Research, Massachusetts Institute of Technology, 70 Vassar St, Cambridge, MA 02139, USA}
\email{fazaid@mit.edu}


\begin{abstract}
We present a high-resolution XRISM/Resolve observation of the persistent black hole candidate 4U~1957+115, revealing the first detection of an intrinsic disk wind in this source. The absorption is dominated by highly ionized Fe~XXV and Fe~XXVI and varies substantially throughout the observation, becoming prominent during the final $\sim40$~ks despite a modest decrease of only $\sim5$--$10\%$ in source flux. The observed evolution of the absorption cannot be explained solely by ionization changes driven by the modest continuum variability, favoring an inhomogeneous outflow with evolving line-of-sight properties. Time-resolved spectroscopy reveals substantially narrower absorption features in shorter intervals, with the inferred turbulent velocity decreasing from $\sim350~{\rm km/s}$ in the longer integration to $\lesssim70~{\rm km/s}$ in the shortest selections. This suggests that the broader absorption profiles measured over longer integrations may result, at least in part, from averaging over an evolving wind rather than from intrinsically larger turbulent velocities. Such time averaging can therefore affect the inferred physical properties of the wind and complicate efforts to identify its launching mechanism. Even though the wind carries only a small fraction of the radiative luminosity in kinetic form, its mass-loss rate may remain dynamically relevant. These results highlight the importance of time-resolved, high-resolution X-ray spectroscopy for characterizing the structure and variability of accretion disk winds.

\end{abstract}



\section{Introduction} \label{sec:intro}

Accretion disk winds are a common feature of luminous black hole (BH) X-ray binaries (XRBs) and provide a direct view of the interaction between the accretion flow and its surrounding environment. In the X-ray band, these outflows are identified as narrow absorption features from highly ionized species, often blueshifted by hundreds to thousands of km/s (\citealt{2016AN....337..368D,2012MNRAS.422L..11P}). Their preferential detection in systems viewed at high inclination indicates that the highly ionized phase of the wind has a predominantly equatorial geometry (\citealt{2012MNRAS.422L..11P}). Additionally, winds are strongly connected to the BH XRB accretion state, with prominent ionized absorption being most commonly observed during bright, disk-dominated soft states, while the features generally weaken or disappear as sources transition toward harder states (\citealt{2009Natur.458..481N, 2012MNRAS.422L..11P}). The inferred mass-loss rates can be comparable to, or even exceed, the mass accretion rate reaching the inner disk, implying that winds may influence the evolution of the accretion flow itself and potentially participate in the observed coupling between accretion states and relativistic jets (\citealt{2009Natur.458..481N, 2012MNRAS.422L..11P}).

Despite their ubiquity, the physical mechanism responsible for launching these winds remains uncertain. Irradiation of the outer disk naturally produces a Compton-heated atmosphere that can become unbound at sufficiently large radii, while radiation pressure can further increase the density and velocity of such thermally driven winds (\citealt{1983ApJ...271...70B, 2018MNRAS.473..838D, 2019MNRAS.490.3098T}). Thermal-radiative calculations can reproduce many of the observed properties of winds in BH XRBs, including their characteristic velocities and their strong dependence on spectral state (\citealt{2019MNRAS.490.3098T}). In contrast, the high densities and small inferred launching radii measured in some systems, most notably GRO J1655-40 and GRS 1915+105, have motivated models in which magnetic pressure or magnetocentrifugal forces play an important role (\citealt{2006Natur.441..953M, 2016ApJ...821L...9M}). The absorbing medium can also be highly dynamic, with changes in line strength, ionization, and velocity being observed from seconds to much longer timescales, reflecting both the response of the plasma to variations in the illuminating continuum and possible changes in the density or geometry of the outflow (\citealt{2018ApJ...860L..19N, 2022ApJ...933..122L}). Resolving this variability is therefore essential to distinguish changes in ionization from genuine structural evolution of the wind and to determine whether the outflow is smooth, stratified, or composed of transient inhomogeneities.

High-resolution X-ray spectroscopy is essential for resolving the narrow absorption features produced by highly ionized disk winds and for measuring their velocity structure, line broadening, and ionization state. The Resolve instrument (\citealt{2025JATIS..11d2023I}) aboard the X-Ray Imaging and Spectroscopy Mission (XRISM; \citealt{2021SPIE11444E..22T}) provides non-dispersive microcalorimeter spectroscopy with an energy resolution of $\lesssim 5$ eV near the Fe K band, enabling velocity and line-profile measurements that are impossible to obtain with conventional CCD spectroscopy. This capability is particularly valuable for BH XRB winds, where Fe XXV and Fe XXVI absorption can be intrinsically narrow and variable, and recent XRISM observations have demonstrated the ability of Resolve to detect and characterize weak absorption features at Fe K energies (\citealt{2025PASJ...77.1210Y}). XRISM therefore provides a new opportunity to follow the evolution of highly ionized outflows on timescales short enough to probe changes in the structure and physical conditions of the wind.

4U 1957+115 is a persistent low-mass X-ray binary (LMXB) and BH candidate that has remained in a disk-dominated soft state throughout its observed history, displaying only weak rapid X-ray variability and a weak hard spectral component (\citealt{2002MNRAS.331...60W, 2008ApJ...689.1199N, 2023ApJ...944..165B}). Within the irradiated disk-instability model, such persistent behavior is expected when the mass-transfer rate is sufficiently high to keep the accretion disk in the hot, ionized state, and 4U~1957+115 has previously been discussed in this context (\citealt{2012MNRAS.424.1991C,2014MNRAS.444.3802H}). The fundamental properties of the binary remain uncertain, largely because the donor star has not been directly detected and therefore no dynamical black hole mass or robust distance measurement is available. Spectral modeling generally favors a rapidly spinning black hole and a relatively high disk inclination (\citealt{2008ApJ...689.1199N, 2023ApJ...944..165B, 2023ApJ...946...19D, 2024A&A...684A..95M}). Despite this combination of a persistently soft spectrum and potentially high inclination, previous high-resolution Chandra and XMM-Newton observations have not revealed a clear intrinsic disk wind, with prior detections of narrow absorption toward the source, including Ne IX and O VII, being attributed to hot gas in the Galactic interstellar medium (\citealt{2008ApJ...672L..21Y, 2008ApJ...689.1199N}). As a result, prior to the XRISM observation presented here, there was no compelling spectroscopic evidence for a highly ionized outflow intrinsic to 4U 1957+115.

In this work, we present a high-resolution XRISM/Resolve observation of 4U 1957+115 and investigate the properties and variability of its highly ionized absorbing material. The paper is structured as follows. In Section \ref{sec:data}, we present the observation and the details of the spectral extraction pipeline used. In Section \ref{sec:analysis} we present the analysis of the Resolve spectra. In particular, in Section \ref{subsec:time-averaged} we offer a view of the time-averaged spectrum, in Section \ref{subsec:spectrogram} we create a spectrogram using the observation, in Section \ref{subsec:line_search} we apply a line-search algorithm to the data, and in Section \ref{subsec:time-resolved} we present a time-resolved analysis of the spectra. Lastly, in Section \ref{sec:results} we summarize our results and put them in context with the current view of the field.

\begin{figure}[t]
    \centering
    \includegraphics[width=\linewidth]{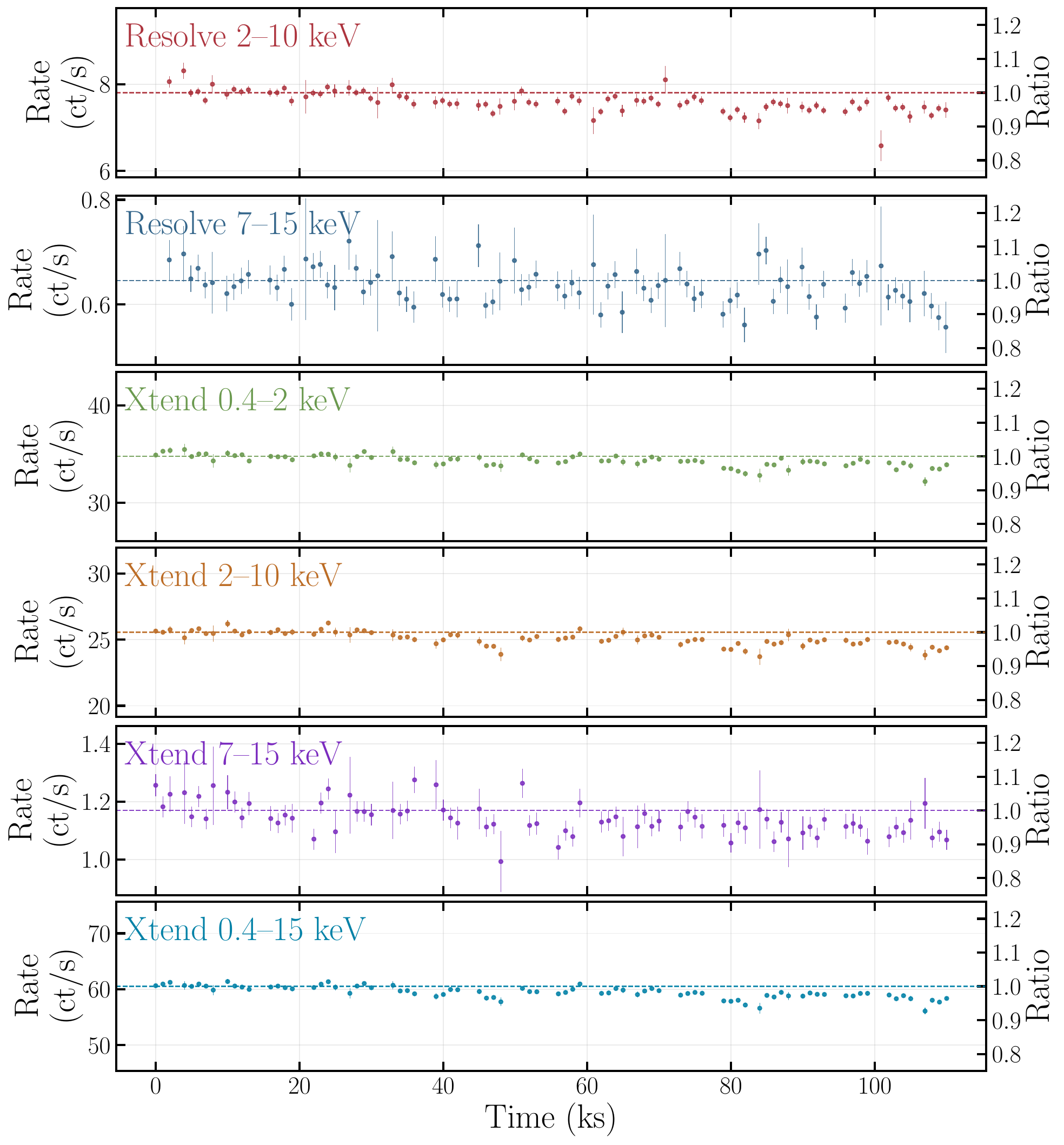}
    \caption{Resolve and Xtend light curves of 4U 1957+115 across multiple energy bands, binned at 1 ks time resolution. The horizontal dashed lines represent the median value of each light curve, calculated over the first 40 ks of the observation (not accounting for dead time). The axes on the right show the ratio between the individual light curves and their corresponding medians.}
    \label{fig:resolve_xtend_light_curves}
\end{figure}

\section{Data}\label{sec:data}

XRISM observed 4U 1957+115 on Nov 24, 2025 under ObsID 202036010 for an effective exposure of 62 ks. 
The data were analyzed using the standard calibration procedures\footnote{\url{https://heasarc.gsfc.nasa.gov/docs/xrism/analysis/index.html}} with the tools available in Heasoft v6.36 and using CALDB version 20260315. The Resolve observation was conducted with the filter wheel open and the Xtend observation was conducted using the 1/8 window mode. The spectral analysis in this work was performed using XSPEC v12.15.1. Figure \ref{fig:resolve_xtend_light_curves} shows the Resolve 2-10 keV, 7-15 keV, and Xtend 0.4-2 keV, 2-10 keV, 7-15 keV, and 0.4-15 keV light curves obtained during the observation, extracted using 1 ks time bins. The horizontal dashed lines represent the medians of the respective light curves during the first 40 ks of spacecraft time, duration chosen arbitrarily to better highlight changes in observed rates. Across data from both instruments and across different energy bands, the source flux decayed by $\sim5-10\%$ throughout the observation. 


The Resolve spectra were extracted from the entire detector, with the exception of pixels 12 (calibration) and 27 (known stability issues), using only high-resolution primary (Hp) events. Throughout the analysis, we used large (L) \texttt{rmf} files, generated using the \texttt{rslmkrmf} command, after excluding low-resolution secondary (Ls) events. For each spectrum extracted throughout the analysis, we generated exposure maps using the \texttt{xaexpmap} command, which were subsequently used to generate \texttt{arf} files through the \texttt{xaarfgen} command. Unless otherwise noted in the text, spectra presented were rebinned using the ``optimal" binning scheme in \texttt{ftgrouppha} (\citealt{2016A&A...587A.151K}). Given the brightness of the source, we did not extract background spectra for Resolve. In the Resolve 2-10 keV band, the source had an average count rate of 7 c/s. The analysis of the Xtend spectra is reserved for a follow-up publication.



\section{Analysis}\label{sec:analysis}
\subsection{Time-averaged analysis} \label{subsec:time-averaged}

\begin{figure*}[t]
    \centering
    \includegraphics[width=0.85\linewidth]{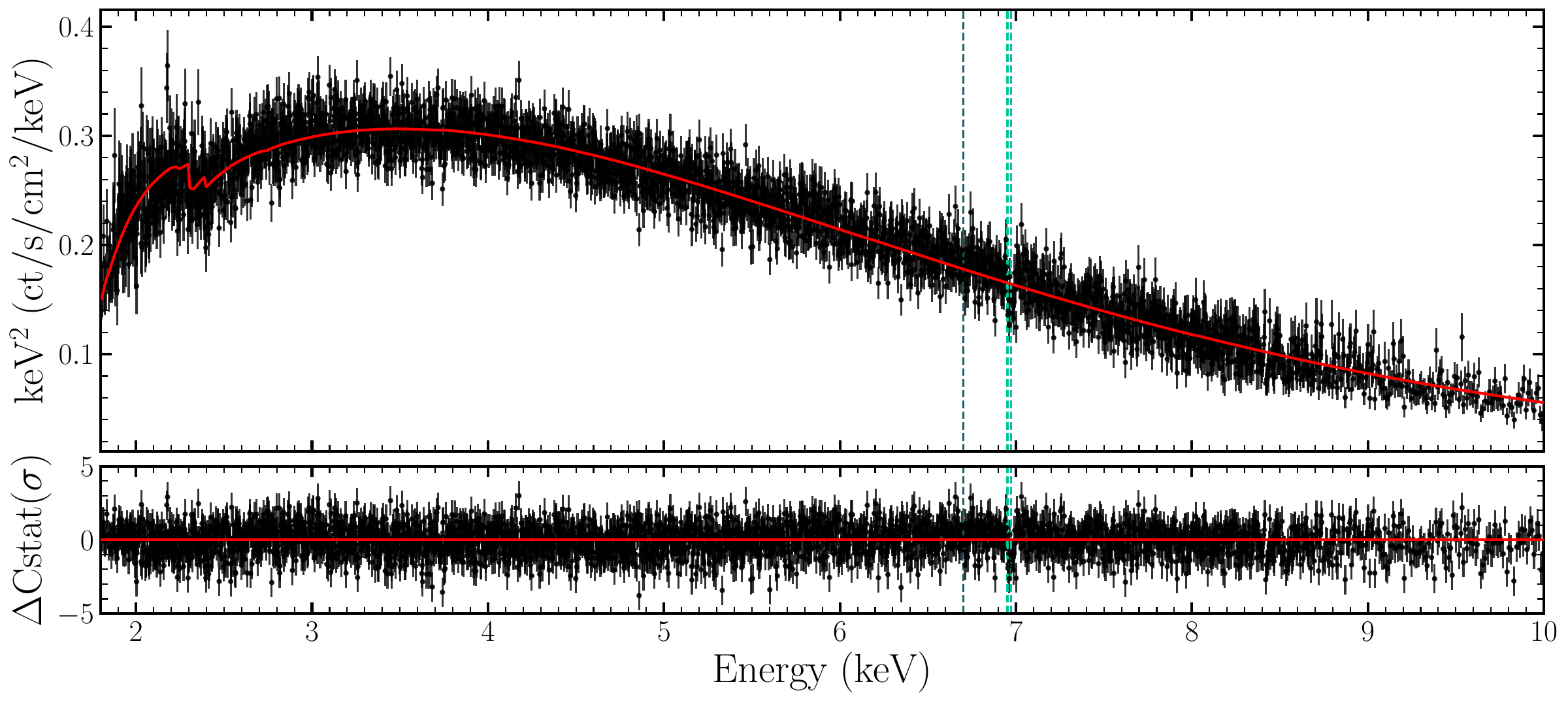}
    \caption{Time-averaged Resolve spectrum of 4U 1957+115. The top panel shows the spectrum (black) and the best-fit continuum model described in Section \ref{subsec:time-averaged}, not accounting for absorption. The bottom panel shows the residuals in terms of Cstat. The vertical dashed lines show the rest energies of Fe XXV and Fe XXVI transitions. }
    \label{fig:spectrum}
\end{figure*}

We extracted a time-averaged Resolve spectrum of the source, together with the L \texttt{rmf} file and the associated \texttt{arf} file. The spectrum is shown in the top panel of Figure \ref{fig:spectrum}, in units of $\rm keV^2\times(ct/s/cm^2/keV)$. An initial fit using the model \texttt{TBabs$\times$diskbb} produces a good fit, with $\rm Cstat/\nu=3037/2879$. The \texttt{diskbb} component accounts for the thermal emission originating from the accretion disk, while the \texttt{TBabs} component models the galactic absorption along the line of sight. However, the residuals of this fit indicate that the model does not fully capture the complexity of the spectrum. Fitting the spectrum with the model \texttt{TBabs$\times$(diskbb+powerlaw)} produces a significantly improved fit to the continuum, with $\rm Cstat/\nu=2889/2877$. The \texttt{powerlaw} component accounts for non-thermal emission from a compact corona. This best-fit continuum model is shown through the red line in Figure \ref{fig:spectrum}, and the bottom panel in Figure \ref{fig:spectrum} shows the residuals of the fit in terms of Cstat. The fit measured the disk temperature to be $kT_{\rm in}\simeq1.46\;\rm keV$, and the power-law index to be $\Gamma\leq4.9$. The column density along the line of sight $N_{\rm H}$ was unconstrained in the fit, taking the lowest allowed value. This is likely due to a combination of limited low-energy coverage and an intrinsically small column along the line of sight, estimated around $10^{21}\;\rm cm^{-2}$ using the HEASARC Column Density tool\footnote{\url{https://heasarc.gsfc.nasa.gov/cgi-bin/Tools/w3nh/w3nh.pl}}.

\begin{figure}[h]
    \centering
    \includegraphics[width=0.95\linewidth]{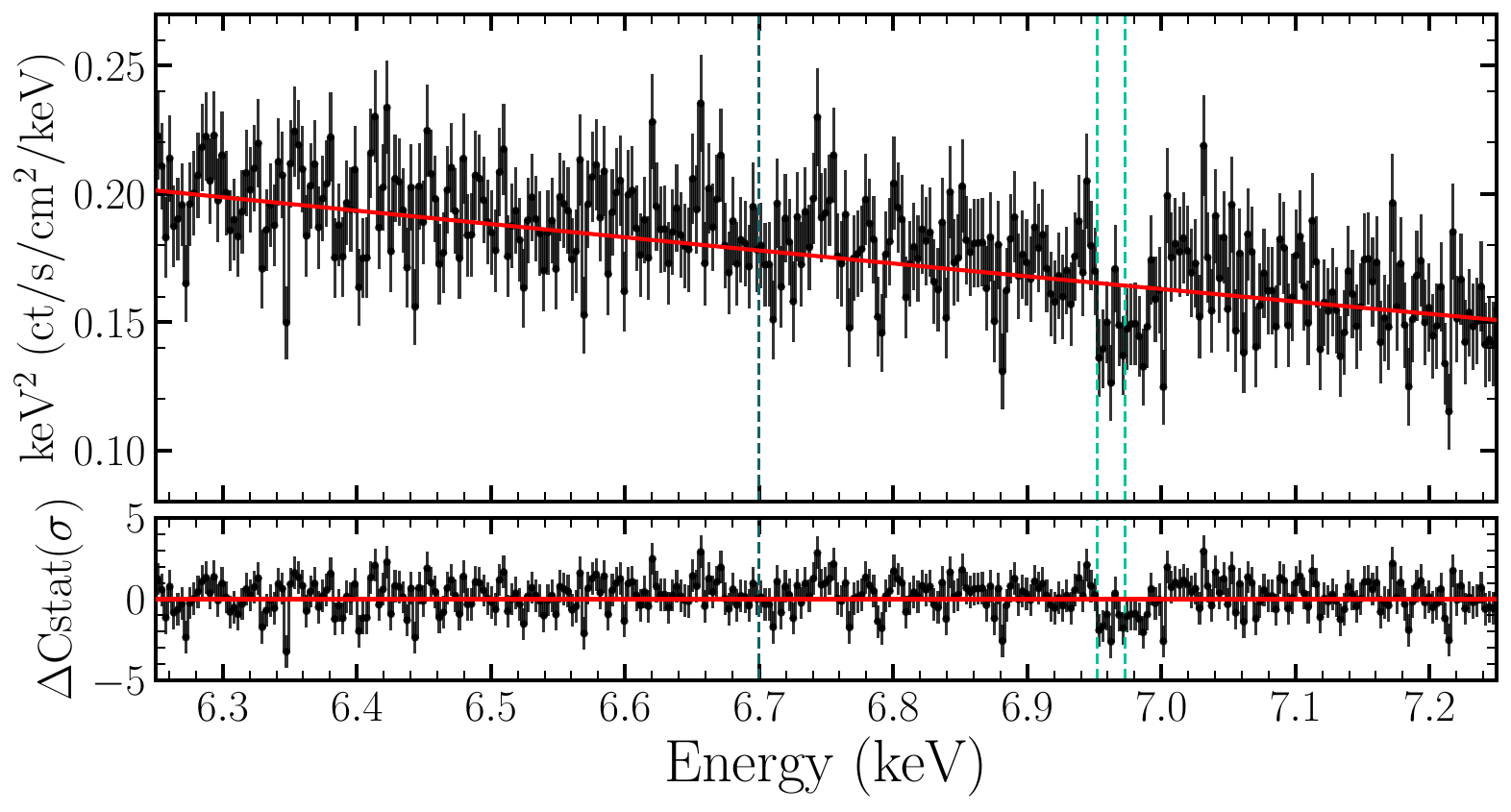}
    \caption{Zoom in on the Fe band of the time-averaged Resolve spectrum of 4U 1957+115. The colors used are the same as in Figure \ref{fig:spectrum}. The absorption feature around 6.97 keV stands out in both the raw data panel and the panel showing the residuals.}
    \label{fig:Fe_band_averaged}
\end{figure}

The most striking feature when visually inspecting the residuals in Figure \ref{fig:spectrum} is an absorption line around 7 keV, consistent with the rest-energy of Fe XXVI. Figure \ref{fig:Fe_band_averaged} shows a zoom-in version, focused around the Fe band (6.25-7.25 keV). Notably, while the absorption structure around 6.97 keV is present, no absorption features are evident around lower-ionization states of Fe. In particular, this time-averaged spectrum shows no evident absorption feature around 6.7 keV (Fe XXV). Furthermore, the spectra show no evidence of narrow Fe K$\alpha$ (at 6.4 keV) or Fe K$\beta$ (at 7.06 keV) emission. Lastly, no strong Fe K edge is present in the spectra (around 7.1 keV), consistent with expectations based on the low galactic column density in the direction of the source. 

Interestingly, the line which we tentatively attribute to Fe XXVI appears to be broadened. To understand whether this apparent broadening is an intrinsic effect or an artifact caused by averaging the spectrum from the entire duration of the observation, we further extracted spectra from fine time slices. This analysis is presented in the following section.

\subsection{Spectrogram}\label{subsec:spectrogram}

To investigate the temporal evolution of spectral features, particularly the Fe-band absorption, we constructed a spectrogram of the Resolve observation. Based on the 2-10 keV light curve, we split the exposure in finely sliced time intervals according to multiple schemes. First, we set an arbitrary duration of 2ks for the spectra. Second, we split the light curve so that each good time interval (GTI) contains 10,000 Hp counts in the 2-10 keV band. Lastly, we performed the same experiment with each spectrum containing 25,000 counts. For clarity, we only present the results of the last of the three experiments, but note that all treatments produce consistent results.

\begin{figure*}[htbp]
    \centering
    \includegraphics[width=0.9\linewidth]{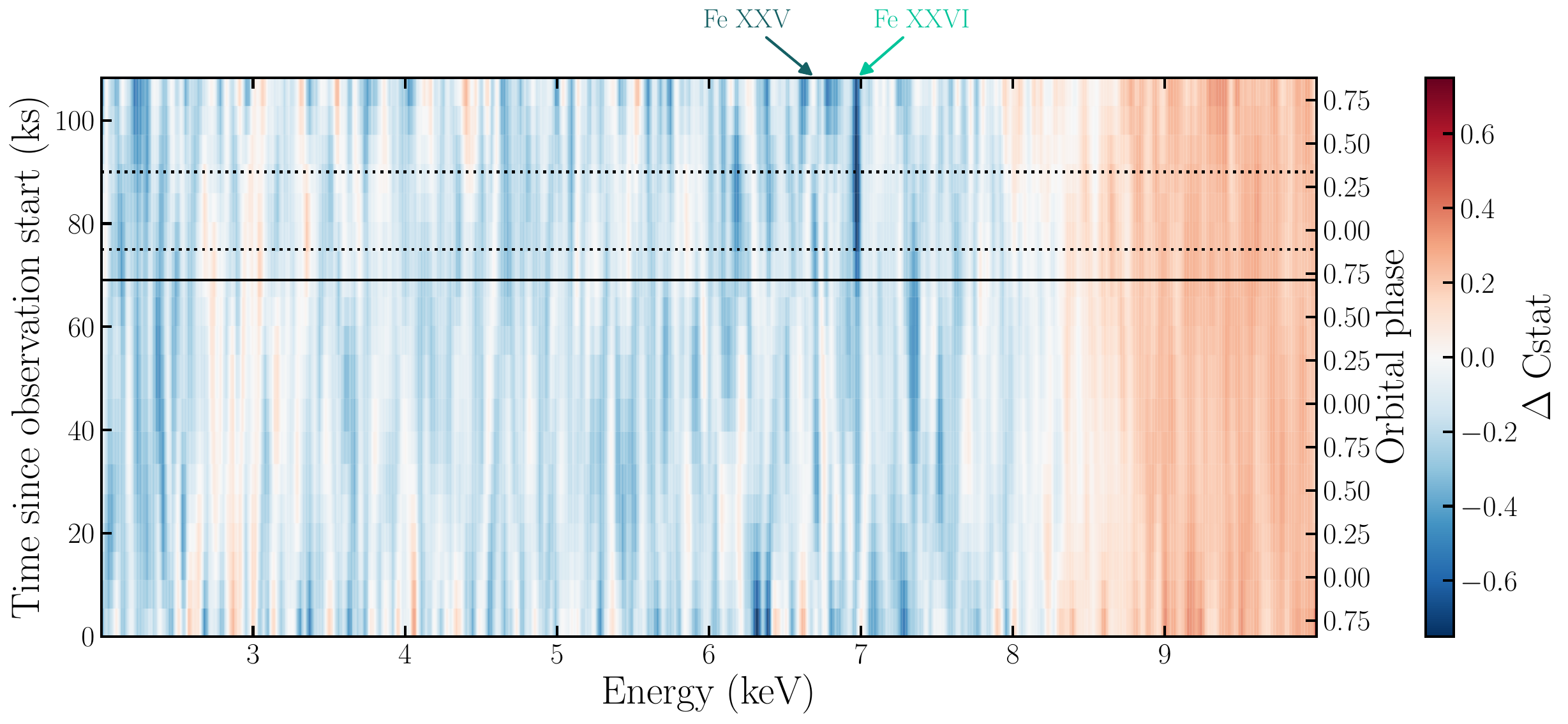}
    \caption{Spectrogram showing contributions to Cstat in the Resolve observation of 4U 1957+115, across the 2-10 keV band. The x-axis represents photon energy, and the y-axis represents time during the observation. The y-axis on the right side of the figure shows the corresponding orbital phase of the system. The residuals are computed with respect to the \texttt{TBabs*(diskbb+powerlaw)} model, with bluer color indicating negative residuals and red colors indicating positive contributions. The duration of each individual time slice was selected to encompass a total of 25,000 Hp counts. The arrows at the top indicate the rest-frame energies of Fe XXV and Fe XXVI. The solid horizontal line shows the point at which the observation was split to extract the ``Part 1" (before) and ``Part 2" (after) spectra (see below). The horizontal dotted lines indicate the interval used for the ``Narrow" spectrum. The spectrogram was smoothened, for visual clarity.}
    \label{fig:full_spectrogram}
\end{figure*}

Splitting the observation into intervals containing 25,000 counts produced 19 GTIs. For each interval, we extracted a spectrum and rebinned it at 3eV resolution. For simplicity, as we did not attempt to measure any fit parameters through this experiment, we used the time-averaged \texttt{rmf} and \texttt{arf} files for analyzing these spectra. We fit all 19 spectra with the same model, \texttt{TBabs$\times$(diskbb+powerlaw)} in the 2-10 keV band and plotted the residuals, normalized by the statistical uncertainty in each spectral bin, as a function of the time elapsed during the observation and the orbital phase of the binary system. The orbital phase was computed according to the ephemeris of \cite{2012AJ....144..108M}: $P=0.388879\;\rm days=9.3331\;\rm hr$ and $T_0=\mathrm{MJD}\;54621.329$, with phase zero corresponding to the peak of the optical flux. The XRISM observation spans roughly three binary orbits. Figure \ref{fig:full_spectrogram} shows this spectrogram in the entire 2-10 keV band, with a 2D Gaussian smoothing applied ($\sigma =1.3$ bins in the temporal direction, and $\sigma =2$ in the spectral direction), for visual clarity. 

The first noticeable element in the spectrogram is an overall excess above 8.5 keV in all fits. This is likely due to the \texttt{powerlaw} component not fully capturing the spectral curvature, and may indicate that a more physically motivated continuum model, including relativistic reflection, could better describe this part of the spectrum. We note, however, that the dynamic range in $\Delta C$-stat in Figure~\ref{fig:full_spectrogram} is small, and the corresponding positive residuals are not statistically significant enough to warrant the addition of a reflection component for the analysis presented here. We discuss this effect further in Section \ref{subsec:time-resolved}.

A second feature to note is narrow, double-peaked absorption between 6.2-6.5 keV in the first $\sim15$ ks of the observation. The proximity of the energies at which these features appear to the neutral Fe K$\alpha$ line at 6.4 keV makes it appealing for further investigation. We extracted a spectrum using only that first segment of the observation to probe this. In the unsmoothed spectra, the features appear as a series of narrow (1-2 spectral bins), low-significance absorption features at 6.27, 6.33, and 6.37 keV. Given how all these are low-significance features, and how in order for the energies they appear at to be caused by redshifted Fe K$\alpha$ their velocities would need to be significantly higher than what could be inferred from their widths, we attribute these to noise in the spectra and do not further investigate their origin. 

\begin{figure*}[htbp!]
    \centering
    \includegraphics[width=0.9\linewidth]{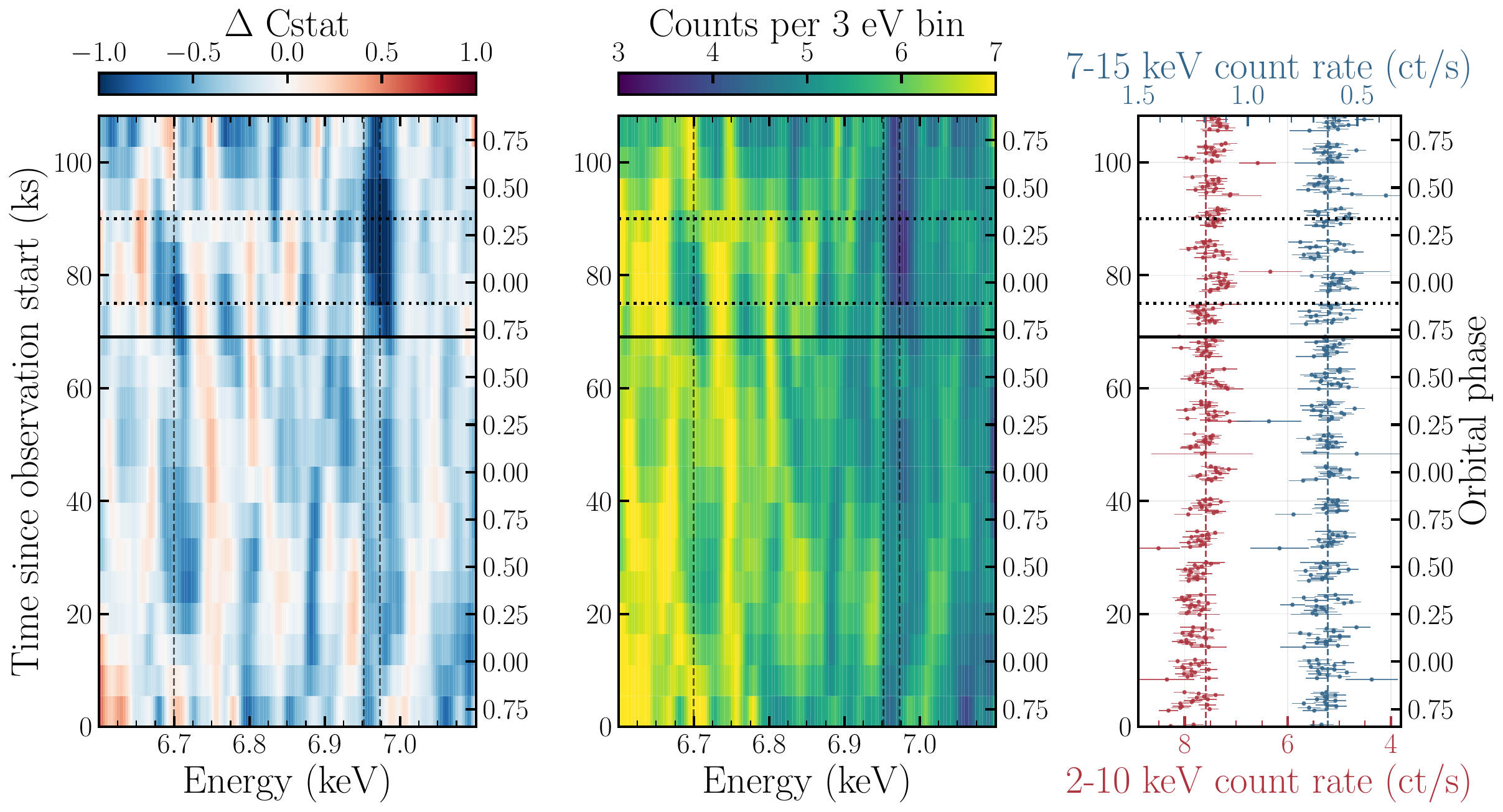}
    \caption{Left: spectrogram of 4U 1957+115, similar to that in Figure \ref{fig:full_spectrogram}, but focused on the 6.6-7.1 keV band, with residuals calculated using a local \texttt{powerlaw} model. Center: Spectrogram showing the number of counts per spectral bin as a function of energy and time during the observation. Both spectrograms were smoothened for visual clarity. Right: Resolve light curves in the 2-10 keV (red) and 7-15 keV (blue) bands, with axes offset to bring the two to the same scale. The vertical dashed lines indicate the median values of the respective light curves. Across the three panels, the horizontal lines have the same meaning as in Figure \ref{fig:full_spectrogram}.}
    \label{fig:Fe_band_spectrogram}
\end{figure*}

The third noticeable effect is that absorption in the Fe band is not persistent, becoming most prominent only $\sim70$ ks after the beginning of the observation. To further assess the temporal structure of absorption in the spectra, we re-fit the 19 segments in the 6.6-7.1 keV band using a simple \texttt{powerlaw} model, under the assumption that locally, the spectrum is relatively flat. The left and central panels in Figure \ref{fig:Fe_band_spectrogram} show the spectrogram in the 6.6-7.1 keV band, focusing on the Fe region. The left panel shows the contribution to Cstat, similarly to Figure \ref{fig:full_spectrogram}. The central panel shows the number of counts per 3 eV spectral bin. The right panel shows the Resolve light curve of the source during the observation, in the 2-10 keV band (red) and 7-15 keV band (blue). The vertical dashed lines show the median value of the light curves in the respective bands.

\begin{figure*}[ht]
    \centering
    \includegraphics[width=0.9\linewidth]{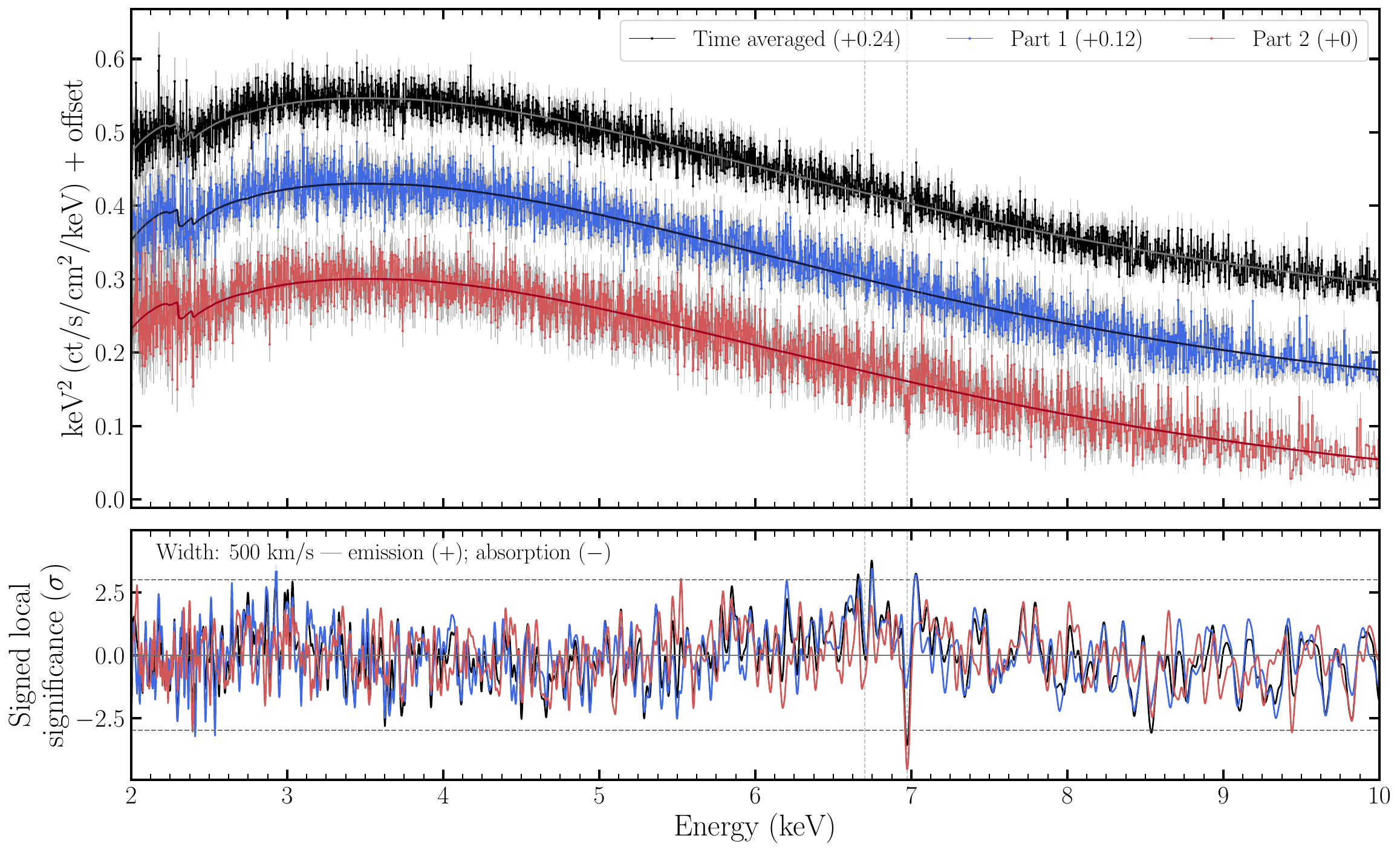}
    \caption{Top: Time-averaged (black), Part 1 (blue) and Part 2 (red) Resolve spectra, offset for visual clarity. Bottom: Local line significance resulting from the line-search algorithm using a velocity width of 500 km/s. The vertical lines show the positions of the Fe XXV and Fe XXVI resonance transitions, and the horizontal dashed lines in the bottom panel show the $\pm3\sigma$ significance thresholds.}
    \label{fig:spectra_and_significance}
\end{figure*}

Figure \ref{fig:Fe_band_spectrogram} further demonstrates the transient nature of absorption during the XRISM observation of 4U 1957+115. Therefore, moving forward, we extracted Resolve spectra from the first 69 ks of the observation (corresponding to 39 ks of effective exposure, hereafter ``Part 1," shown in blue), and the remaining 39 ks segment (23 ks of effective exposure, hereafter ``Part 2," shown in red), after the onset of the absorption features. We identify the time where the two segments were separated by the solid horizontal lines in Figures \ref{fig:full_spectrogram} and \ref{fig:Fe_band_spectrogram}. To further investigate the temporal evolution of the absorption features, we also considered shorter, 15~ks intervals throughout the observation. The strongest Fe-band absorption was identified between 75 and 90~ks, corresponding to 8.4~ks of effective exposure (indicated by the horizontal dotted lines in Figures \ref{fig:full_spectrogram} and \ref{fig:Fe_band_spectrogram}). We refer to the spectrum extracted from this interval as the ``Narrow'' spectrum and analyze its properties in Section \ref{subsec:time-resolved}. For reference, the entire observation spanned 108 ks, with 62 ks of effective exposure. These spectra were also rebinned using the ``optimal" binning scheme in \texttt{ftgrouppha}. The top panel of Figure \ref{fig:spectra_and_significance} shows the time-averaged spectrum (black) and the two parts of the observation, displaced vertically for visual clarity. 

\subsection{Line search algorithm}\label{subsec:line_search}

To identify narrow absorption and emission features in the spectra, we ran a line-search algorithm on the time-averaged spectrum and on the two segments (first 69ks and last 39ks). The line search works as follows. We first fit the continuum of the spectra using the model presented above, \texttt{TBabs$\times$(diskbb+powerlaw)}. After fixing the parameters of the continuum model, we added a \texttt{gaussian} component to the model at fixed energies, with fixed width, and fit for the normalization of the component. The normalization was allowed to take both negative and positive values, to explore the presence of narrow absorption and emission features. After fitting and assessing the change in fit statistic, the algorithm changes the energy of the line, and repeats the process. For each spectrum, we ran the scan using four different line widths, corresponding to velocities of 50 km/s, 200 km/s, 500 km/s, and 1000 km/s. The search was performed with a resolution of 2 eV (the size of the step in energy) between 2-10 keV. 

Table \ref{tab:single-line-search} in Appendix \ref{sec:line_search_results} shows the resulting significance in terms of $\sigma$ of any absorption or emission features detected in any of the three spectra. The bottom panel in Figure \ref{fig:spectra_and_significance} shows the local significance derived by the search using a width of 500 km/s. Note that this significance is only local, it does not account for the look-elsewhere effect, and assumes a single free parameter (line normalization). There are few instances of significant ($\geq3\;\sigma$) detections, most appearing as absorption lines. The most significant detection (at $>4.5\sigma$) is the absorption feature at 6.974 keV in the last segment of the observation, that is consistent with Fe XXVI, as identified in Section \ref{subsec:time-averaged}. 

Additionally, an emission feature is detected across most of the velocity scans at 6.746 keV in the time-averaged spectrum and in the first segment of the observation. While this feature can be interpreted as mildly blue-shifted Fe XXV emission, it can also be interpreted as a local mischaracterization of the continuum, with the spectrogram in Figure \ref{fig:Fe_band_spectrogram} showing a mild reduction in counts (and fit statistic) around 6.7 keV, followed by an increase in flux around 6.75 keV. This effect is noticeable in the bottom panel of Figure \ref{fig:spectra_and_significance}, where the general significance trend (i.e., underlying residuals of the continuum) indicates a broad deviation from the zero line. This trend resembles the residuals expected to be produced by unaccounted relativistic reflection. We tested whether introducing a reflection model to our continuum impacts the overall fit and the inferred line properties by replacing the \texttt{powerlaw} component in our best-fit model with the \texttt{relxilllp} variant of the \texttt{relxill} family of models (\citealt{2014ApJ...782...76G, 2014MNRAS.444L.100D}). For the Part 2 spectrum, where the detection of the wind component was strongest, replacing the \texttt{powerlaw} component ($\rm Cstat/\nu=2362.6/2471$) with \texttt{relxilllp} ($\rm Cstat/\nu=2346.3/2464$) decreases the fit statistic by $\Delta \rm Cstat=16.3$ for 7 extra free parameters, representing a $2\sigma$ improvement. However, this improvement corresponds to a decrease of 2.3 in the Akaike Information Criterion (AIC; \citealt{1974ITAC...19..716A}) and an increase of 38.6 in the Bayesian Information Criterion (BIC; \citealt{1978AnSta...6..461S}). The negligible decrease in AIC and the increase in BIC suggest that a reflection component is not statistically significant for this dataset. Nevertheless, we further tested the impact of including relativistic reflection to the model on the properties of the absorption component (see Section \ref{subsec:time-resolved}). We found that all inferred properties of the absorption component are fully consistent, regardless of the choice of the continuum model. Therefore, we continue our analysis using the \texttt{powerlaw} component in our continuum model, but acknowledge that the significance levels of the narrow features reported in this section may be slightly impacted by a different continuum choice. 

At lower energies, the most significant tentative detections are an absorption feature only in the first part of the observation at 2.538 keV (at 4$\sigma$ for broadening under 200 km/s) and an emission feature at 2.930 keV, again identified only in the first part of the observation, favoring broadening $\geq200$ km/s.  It is important to note that these significance calculations do not account for the look-elsewhere effect. Given the weak significance of these detections, we do not investigate the low-energy lines further, but note their tentative detection.

While scanning for a single narrow feature in the spectra can provide initial insights into possibly interesting detections, in the case of such low-significance measurements, a more physically-motivated approach could strengthen the inference. We performed a second scan of the spectra, focusing on identifying features that resemble the Fe XXV and Fe XXVI complex. The choice to focus on the two most highly ionized ions of Fe was made based on the presence of Fe XXVI absorption in the spectra, but the lack of clear signs of lower-ionization states. In this updated search, on top of the continuum, we added three \texttt{zgauss} components with energies fixed at the rest energies of the Fe XXV He$\alpha$ and the Fe XXVI Ly$\alpha$1 and Ly$\alpha$2 resonance lines, and linked redshift parameter. Additionally, we linked the widths of the three lines, and set the normalization of the Ly$\alpha$1 line (at 6.973 keV) to be twice that of the Ly$\alpha$2 line (at 6.952 keV), while the normalization of the Fe XXV line was allowed to vary independently. We tested the same 4 broadening velocities, and varied the redshift of the entire complex of 3 lines to scan the 6-8 keV band in increments of 2 eV. 

Table \ref{tab:three-line-search} in Appendix \ref{sec:line_search_results} shows all trials that produced an improvement in fit statistic of $\geq1.5\;\sigma$. The only significant detections $\geq3\;\sigma$ were obtained at $z=-8.9\times10^{-4}$ ($\sim-270\;\rm km/s$) and $z=-0.039$ ($\sim-12000\;\rm km/s$). However, the second solution is highly unlikely, with the fit identifying the 6.97 keV Fe XXVI lines as a blue-shifted version of the 6.7 keV Fe XXV line, and adjusting the normalization of its corresponding Fe XXVI lines to vanish. This experiment suggests that the spectra favor a single, slightly blue-shifted absorption layer as opposed to multiple absorption layers with different speeds and broadening.

\subsection{Time- and phase-resolved analysis}\label{subsec:time-resolved}
Following the results of the line search, particularly the three-line experiment, we fit the Resolve spectra using the self-consistent photoionization model \texttt{pion} (\citealt{2015Natur.526..542M, 2016A&A...596A..65M}). As \texttt{pion} is currently only available within \texttt{SPEX}, we used the ``SPEX-to-XSPEC" \footnote{\url{https://github.com/yrxu/Tools-in-XSPEC}} package (\citealt{2019MNRAS.490..683P, 2024A&A...687A.179X}) to generate \texttt{xspec}-compatible multiplicative tables. We first fit the continuum of the source in \texttt{SPEX} and used it to compute \texttt{pion} models using an array of 10 linearly-spaced points in ionization between $\log\xi=3-7$, 10 logarithmically-spaced points in column density between $\rm N=10^{21}\;cm^{-2}$ and $\rm N=10^{25}\;cm^{-2}$, and 10 logarithmically-spaced points in turbulent velocity between $v_{\rm turb}=1-10^4\;\rm km/s$. The ranges of the parameters were chosen based on preliminary fits using broader ranges and coarser spacing. We note that the use of precomputed \texttt{pion} tables in \texttt{xspec} does not allow the ionizing spectral energy distribution to respond self-consistently to changes in the fitted continuum parameters, as would be possible when fitting the continuum and photoionized absorber simultaneously in \texttt{SPEX}. Consequently, the inferred absorber properties depend on the assumed ionizing continuum, introducing an additional source of systematic uncertainty, particularly for the column density and ionization parameter.

We modified the previously obtained best-fit models to include the multiplicative \texttt{pion} component, producing the following model: \texttt{TBabs $\times$ pion $\times$ (diskbb + powerlaw)}. First, we fit the Part 2 spectrum, obtaining a good fit, with $\rm Cstat/\nu=2362.68/2471$. This represents a $4\sigma$ detection when compared to the $\rm Cstat/\nu=2387.41/2475$ obtained without the \texttt{pion} component. The \texttt{pion} component measures a high ionization $\log\xi=5.8_{-0.5}^{+0.3}$, a column density $N_{\rm H,\; PION}=4.8_{-3.7}^{+1.7}\times10^{23}\;\rm cm^{-2}$, a turbulent velocity $v_{\rm turb}=350_{-170}^{+290}\;\rm km/s$, and a blueshift of $z=1.4_{-0.6}^{+0.4}\times10^{-3}$ ($420_{-190}^{+150}\;\rm km/s$), when fixing the covering fraction to 1. The uncertainties on the parameters were assessed through an MCMC analysis on the best-fit model for the Part 2 spectrum. The corner plot resulting from the analysis is shown in Figure \ref{fig:corner} in Appendix \ref{sec:corner_plots}. However, it is extremely important to note that the MCMC analysis reveals a strong, nonlinear correlation between the ionization parameter and the column density of the absorbing material. At higher ionization, a broad range of column densities can reproduce the observed absorption, while lower-ionization solutions occupy a much narrower region of parameter space. Indeed, we identify an alternative solution with lower ionization ($\log\xi=4.8$) and column density ($N_{\rm H}=4\times10^{22}~{\rm cm^{-2}}$), comparable to those measured in the shorter integrations (see below), that provides a virtually identical fit statistic ($\Delta C_{\rm stat}=0.05$). The relative density of MCMC samples across these regions may be influenced by the geometry of the parameter space and the efficiency with which it is explored, and should not be interpreted as evidence that the higher-ionization solution is uniquely preferred.  Importantly, the turbulent and bulk outflow velocities inferred from this alternative solution remain consistent with those obtained from the original fit. Consequently, while the large column density inferred for Part~2 is not uniquely required by the data, the inferred kinematic properties appear less sensitive to this degeneracy, and physical quantities derived from the ionization and column density should be interpreted with caution.

The 6.6-7.1 keV band of the spectrum and the resulting model are shown in red in Figure \ref{fig:Fe_band_resolved}. Figure \ref{fig:full_spec_parts1and2} in Appendix \ref{sec:spectra_plots} shows the full 2-10 keV spectra for both parts of the observation. Additionally, Figure \ref{fig:time_split_wide} in Appendix \ref{sec:spectra_plots} shows the 6.4-10 keV bands, and the rest-energies of strong transitions of Fe XXV, Fe XXVI, Ni XXVII, and Ni XXVIII.  

\begin{figure}
    \centering
    \includegraphics[width=0.95\linewidth]{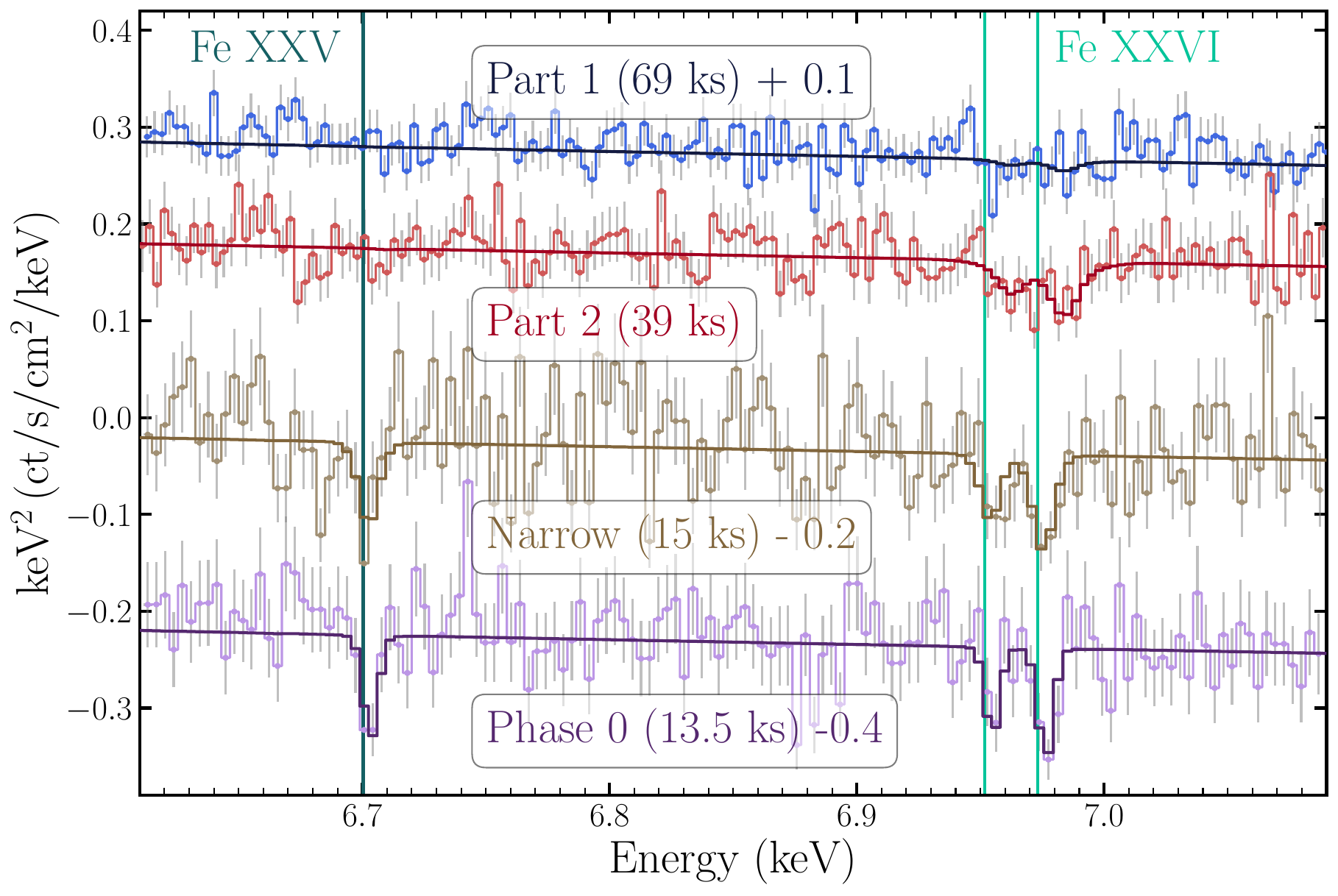}
    \caption{Time-resolved spectra obtained from the Resolve observation of 4U 1957+115, focusing on the Fe band (6.6-7.1 keV). The ``Part 1" and ``Part 2" spectra (blue and red) were obtained from the first 69 ks (39 ks of effective exposure) and last 39 ks (23 ks of effective exposure), respectively. The ``Narrow" spectrum (brown) was obtained from the 75-90 ks interval during the observation (8.4 ks of effective exposure). The ``Phase 0" spectrum (purple) was obtained by phase-folding the observation and extracting the spectrum from a phase bin of width of 0.125 centered at 0 (effective exposure of 8.5 ks). The numbers on the y axis correspond to the Part 2 spectrum, with the other 3 spectra being offset for visual clarity. The vertical green lines indicate the rest energies of the Fe XXV and Fe XXVI transitions. The solid curves represent the best-fit models described in the paper. }
    \label{fig:Fe_band_resolved}
\end{figure}

To highlight the change in absorption structure, we used the best-fit model for Part 2 of the observation to fit Part 1 (shown in blue in Figure \ref{fig:Fe_band_resolved}). Note that the addition of the \texttt{pion} component to the Part 1 spectrum only improves the fit at $\sim0.7\sigma$ level, and cannot be regarded as a statistically significant detection. Nevertheless, we tested four scenarios. First, we tested whether the change in absorption features was caused by a change in ionization of the material. We fixed all the parameters in the \texttt{pion} component to the ones obtained using the spectrum from Part 2, but allowed the ionization to vary freely, measuring $\log\xi\geq6.4$ (at $1\sigma$ confidence assessed through the \texttt{xspec} \texttt{error} function). Given the apparent lack of absorption in the Part 1 spectrum, this value would translate to an over-ionization of the obscuring material. However, comparing the lower limit of $\log\xi\geq6.4$ in Part 1 with the upper limit of the $1\sigma$ confidence limit on the ionization for Part 2 of 5.8+0.3=6.1, the difference in ionization parameter for equal column density translates to a required decrease in ionizing luminosity by a factor of greater than 2 when assuming a constant density. This scenario is excluded because of the small reduction in observed flux between the two parts of the observation, on the order of 5$\%$ (see Figure \ref{fig:resolve_xtend_light_curves}).

Second, we tested whether the change can be attributed to an increase in covering fraction of the obscurer as the time in the observation progresses. Fixing all parameters in the model to the value obtained when fitting the spectrum from Part 2 except for the covering fraction (assumed to be 1 in the fit to the Part 2 spectrum) measures a covering fraction of $\leq0.4$ for the first part of the observation. Third, we tested whether the change in absorption can be attributed to a change in column density of the obscurer. Applying the same treatment but allowing only the column density to vary, we measure $N_{\rm H}\leq1.3\times10^{22}\;\rm cm^{-2}$ ($\leq1\times10^{22}\;\rm cm^{-2}$ for the alternative Part 2 solution). Lastly, we allowed all parameters of the \texttt{pion} model to vary. However, the weak absorption lines did not allow a significant improvement in fit statistic over any of the three fits described above. The first three scenarios produced $\rm Cstat/\nu\simeq2690/2653$, with the fourth one producing $\rm Cstat/\nu\simeq2688/2650$. 
All four scenarios provide statistically comparable fits, indicating that the weak or absent absorption in Part~1 can be explained by changes in ionization, covering fraction, or column density, or by a combination of these effects. However, because the absorption features are not significantly detected in Part~1, the data only constrain the range of parameter values consistent with their absence, rather than providing robust measurements of the absorber properties.

We next considered the ``Narrow'' spectrum introduced in Section \ref{subsec:spectrogram}, extracted from the 75--90~ks interval. This spectrum is shown in Figures \ref{fig:Fe_band_resolved} and \ref{fig:time_split_wide} in brown (indicated by the label ``Narrow"). Interestingly, focusing on the Fe band in Figure \ref{fig:Fe_band_resolved} shows that the Fe XXV and Fe XXVI lines are narrower compared to the Part 2 spectrum, and still moderately blueshifted. Notably, the Fe XXV line around 6.7 keV stands out more significantly from the continuum compared to the Part 2 spectrum. It is important to acknowledge that the ``Narrow" spectrum represents a temporal subset of the ``Part 2" spectrum. 

Applying the same model described previously to this ``Narrow" spectrum measures a significantly lower ionization $\log\xi=4.5_{-0.3}^{+0.4}$, which, in turn, leads to a lower column density $N_{\rm H}=4_{-2}^{+7}\times10^{22}\;\rm cm^{-2}$ through the correlation between the two parameters highlighted in the corner plot in Figure \ref{fig:corner}. As visually suggested by the widths of the lines in Figure \ref{fig:Fe_band_resolved}, the measured turbulent velocity of the \texttt{pion} component is lower, but still consistent within $1\sigma$, compared to that of Part 2: $v_{\rm turb}=120_{-80}^{+120}\;\rm km/s$. Lastly, the measured blueshift of the component is $z=4\pm2\times10^{-4}$ ($120_{-60}^{+70}\;\rm km/s$), significantly lower compared to that measured in the Part 2 spectrum ($420_{-190}^{+150}\;\rm km/s$). The lower bulk velocity of the absorption component and the reduced width compared to the one inferred from the Part 2 spectrum may suggest that, in addition to a change in column density and potentially ionization, the obscurer is experiencing significant acceleration throughout the duration of the observation, either in a linear or radial direction (i.e. orbital motion). 

We also extracted spectra from the intervals immediately preceding (69--75~ks) and following (90--105~ks) the Narrow interval to investigate the evolution of the absorption. However, the addition of the wind component improves the fits at only $\sim1\sigma$ and $\lesssim2\sigma$, respectively. These spectra therefore do not provide sufficient constraints to establish whether the outflow velocity evolves continuously or whether the observed differences arise from distinct absorbing structures.

To further test this hypothesis, despite the lack of clear evidence for orbital modulation in the spectrograms in Figures \ref{fig:full_spectrogram} and \ref{fig:Fe_band_spectrogram}, we folded the light curve of the observation on the known 9.3331 hr orbital period and extracted spectra from 8 phase bins, with each spectrum covering a phase interval of $\Delta\phi=0.125$. Of the 8 phase bins, a \texttt{pion} component was only statistically required for the spectrum corresponding to a phase width of $\pm0.062$ around phase 0 (superior conjunction - BH closer to the observer). The resulting spectrum from this phase bin (hereafter ``Phase 0") is shown in purple in Figures \ref{fig:Fe_band_resolved} and \ref{fig:time_split_wide}. For this spectrum, adding the \texttt{pion} component to the model improves the fit statistic by $\Delta \rm Cstat=41$ for 4 additional free parameters, corresponding to a detection at the $5.6\sigma$ level. Notably, when focusing on this narrow band in orbital phase, despite the short 8.5 ks effective exposure within the spectrum, both the Fe XXV and the Fe XXVI doublet appear even narrower compared to either of the previous approaches at dividing the observation. For this spectrum, the \texttt{pion} model measures an ionization consistent with that in the ``Narrow" spectrum,  $\log\xi=4.5\pm0.3$, column density $N_{\rm H}=6_{-3}^{+12}\times10^{22}\;\rm cm^{-2}$, and blueshift, $z=4.6_{-0.8}^{+1.1}\times10^{-4}$ ($140_{-20}^{+30}\;\rm km/s$). However, the inferred turbulent velocity of the \texttt{pion} component is even lower than that of the Narrow spectrum, $v_{\rm turb}\leq70\;\rm km/s$, but still formally within $1 \sigma$. 

The best-fit photoionization parameters obtained from the Part 2, Narrow, and Phase 0 spectra are summarized in Table \ref{tab:wind_parameters}.

\begin{deluxetable}{l|ccc}
\tablecaption{Photoionization parameters measured from the time- and phase-resolved spectra.
\label{tab:wind_parameters}}
\tablehead{
\colhead{Parameter} &
\colhead{Part 2} &
\colhead{Narrow} &
\colhead{Phase 0}
}
\startdata
Exposure (ks)
    & $23$ & $8.4$ & $8.5$ \\
$\log\xi$
    & $5.8^{+0.3}_{-0.5}$
    & $4.5^{+0.4}_{-0.3}$
    & $4.5\pm0.3$ \\
$N_{\rm H}\;(10^{22}\,{\rm cm^{-2}})$
    & $48^{+17}_{-37}$
    & $4^{+7}_{-2}$
    & $6^{+12}_{-3}$ \\
$v_{\rm turb}\;({\rm km/s})$
    & $350^{+290}_{-170}$
    & $120^{+120}_{-80}$
    & $<70$ \\
$v_{\rm out}\;({\rm km/s})$
    & $420^{+150}_{-190}$
    & $120^{+70}_{-60}$
    & $140^{+30}_{-20}$ \\
\enddata
\tablecomments{Exposure refers to the effective exposure time. Uncertainties are reported at the $1\sigma$ confidence level. The turbulent velocities and outflow velocities are those inferred from the \texttt{pion} component.  For Part~2, an alternative solution with $\log\xi=4.8$ and $N_{\rm H}=4\times10^{22}~{\rm cm^{-2}}$ provides a virtually identical fit statistic ($\Delta C_{\rm stat}=0.05$). The strong correlation between ionization and column density therefore introduces additional uncertainty not fully represented by the quoted parameter intervals.}
\end{deluxetable}

\section{Results and Discussion}\label{sec:results}

The main results of the XRISM/Resolve observation of 4U~1957+115 can be summarized as follows:

\begin{itemize}

    \item We detect the first compelling spectroscopic evidence for an intrinsic disk wind in 4U~1957+115. During the latter part of the observation, photoionization modeling measures a highly ionized absorber with $\log\xi=5.8^{+0.3}_{-0.5}$, $N_{\rm H}=4.8^{+1.7}_{-3.7}\times10^{23}~{\rm cm^{-2}}$, and an outflow velocity of $420^{+150}_{-190}~{\rm km/s}$.

    \item The wind is strongly variable. Fe-band absorption is weak or absent during the first part of the observation and becomes prominent only after $\sim70$~ks. The small contemporaneous change in continuum flux cannot account for this behavior through ionization alone, favoring changes in the amount or geometry of absorbing material and therefore an inhomogeneous, likely clumpy wind.

    
    \item The absorption features become narrower in shorter integrations, indicating that unresolved variability in the wind's velocity structure likely contributes to the apparent line broadening measured in longer integrations, rather than this broadening arising entirely from intrinsic turbulence.

    \item Phase-resolved spectroscopy shows the strongest and narrowest absorption near orbital phase zero, where the \texttt{pion} component is detected at $5.6\sigma$ and $v_{\rm turb}\lesssim70~{\rm km/s}$. However, because the observation samples only approximately three binary orbits, we do not interpret this as evidence for a strictly phase-locked wind.

\end{itemize}

The detection of a disk wind in 4U~1957+115 is notable given that previous high-resolution observations of the source found narrow absorption associated with the Galactic interstellar medium but no compelling evidence for an intrinsic outflow (\citealt{2008ApJ...672L..21Y,2008ApJ...689.1199N}). The highly ionized Fe XXV and Fe XXVI absorption detected with Resolve, together with its measurable blueshift and strong temporal variability, instead clearly associates the absorbing material with the binary. The absorption component is detected at $4\sigma$ level for the ``Part 2" spectrum, $4.9\sigma$ for the ``Narrow" spectrum, and $5.6\sigma$ for the ``Phase 0" spectrum. These levels of significance were estimated using an asymptotic likelihood-ratio approximation, treating the improvement in $\rm Cstat$ as following a $\chi^2$ distribution with 4 degrees of freedom. Because the absorber parameters are not fully identifiable under the null hypothesis of no absorption, this significance should be interpreted with caution.

The rapid appearance and disappearance of the Fe-band absorption provides important information about the structure of the wind. Similar short-timescale variability has been observed in other BH XRB disk winds. In H1743--322, Fe XXV absorption was found to vary on timescales of a few hundred seconds, with the observed changes requiring a combination of ionization and geometric effects and allowing for an interpretation in terms of a clumpy outflow (\citealt{2006ApJ...646..394M}). Conversely, flux-resolved spectroscopy of GRS~1915+105 demonstrated that sufficiently large changes in the illuminating continuum can produce substantial wind variability through progressive over-ionization (\citealt{2012MNRAS.421..502N}). We tested whether the variability could be explained by changes in ionization driven solely by the observed continuum evolution. Suppressing the absorption in Part~1 requires $\log\xi\gtrsim6.4$, compared with $\log\xi=5.8^{+0.3}_{-0.5}$ in Part~2, implying a change in ionizing luminosity substantially larger than observed if the density and location of the absorber remain unchanged. However, because $\xi=L_{\rm ion}/nr^2$, variations in the density or location of the absorbing material could also change its ionization state without requiring substantial continuum variability. The rapid appearance of strong Fe-band absorption therefore favors an inhomogeneous wind with evolving line-of-sight properties, potentially involving discrete clumps or a more continuously stratified outflow (\citealt{2006ApJ...646..394M,2016ApJ...821L...9M}). The present observations do not uniquely distinguish between these geometries.

The inclination of 4U~1957+115 provides an additional constraint on the geometry of the wind. Highly ionized Fe-band absorption is detected preferentially in high-inclination BH XRBs, which has motivated a predominantly equatorial wind geometry (\citealt{2012MNRAS.422L..11P}). Although the inclination of 4U~1957+115 remains uncertain, recent spectral and polarimetric studies generally favor a moderate to high inclination ($50^\circ\lesssim\theta\lesssim80^\circ$; \citealt{2008ApJ...689.1199N, 2023ApJ...944..165B, 2023ApJ...946...19D, 2024A&A...684A..95M}). The XRISM detection is therefore broadly consistent with an equatorial wind geometry. At the same time, if the true inclination lies toward the lower end of the allowed range ($\gtrsim50^\circ$), the detection would imply that the highly ionized wind has an appreciable vertical extent, or that clumpy structures can extend above the densest equatorial layers and intermittently cross the line of sight.

The progressive narrowing of the absorption features as the integration time is reduced also has implications for how line widths measured in BH XRB winds should be interpreted. During the 75--90~ks ``Narrow'' interval, the inferred turbulent velocity decreases to $v_{\rm turb}=120^{+120}_{-80}~{\rm km/s}$, compared with $350^{+290}_{-170}~{\rm km/s}$ in the longer Part~2 spectrum, while the measured bulk outflow velocity also decreases from $420^{+150}_{-190}$ to $120^{+70}_{-60}~{\rm km/s}$. This suggests that at least part of the apparent broadening measured in longer integrations may arise from unresolved temporal evolution of the absorber velocity, rather than from intrinsic turbulence alone. High-resolution studies of other BH XRB winds have similarly shown multiple velocity components and time-dependent line centroids that can blend into broader features when averaged over longer intervals (\citealt{2016ApJ...821L...9M,2016AN....337..368D}). The Resolve spectra of 4U~1957+115 therefore highlight how time-averaged line widths may encode both microscopic broadening and macroscopic variability in the outflow.

For comparison, the expected thermal broadening of Fe ions at $T=10^7$~K corresponds to a Gaussian velocity dispersion of $\sigma_{\rm th}\simeq39~{\rm km/s}$, or a full width at half maximum of approximately $90~{\rm km/s}$. The upper limit of $v_{\rm turb}\lesssim70~{\rm km/s}$ measured in the Phase~0 spectrum therefore allows only relatively modest additional turbulent broadening. This suggests that the narrowest absorption features may be close to the thermal broadening regime, while the larger widths measured in longer integrations may partly reflect unresolved bulk velocity variations rather than intrinsic turbulence.

To investigate whether orbital geometry contributes to this kinematic variability, we examined the spectra as a function of binary phase. The absorber is statistically required only in the $\Delta\phi=0.125$ interval centered on phase~0, which corresponds approximately to superior conjunction of the donor star, with the compact object between the donor and the observer (\citealt{2012AJ....144..108M}). In this phase interval, the \texttt{pion} component is detected at $5.6\sigma$ and the absorption features are especially narrow, with $v_{\rm turb}\lesssim70~{\rm km/s}$ and $v_{\rm out}=140^{+30}_{-20}~{\rm km/s}$. If this concentration of absorption at a particular orbital phase is persistent, it would imply that the absorbing material is not azimuthally symmetric around the disk, but instead occupies a preferred region in the binary frame. Such a geometry could arise from a vertically extended structure associated with the stream--disk interaction, or from an azimuthally structured disk wind in which enhanced-density material reaches the line of sight only over a limited range of orbital phase (\citealt{1996ApJ...470.1024A, 2018MNRAS.473..838D}). Phase-dependent velocity shifts of order $\sim100~{\rm km/s}$ have also been reported in the disk wind of GRO~J1655--40, demonstrating that orbital motion can imprint measurable Doppler shifts on wind absorption lines (\citealt{2012MNRAS.421.3550Z}). However, the XRISM observation of 4U~1957+115 samples only approximately three binary orbits, and the spectrogram does not reveal an obvious repeating pattern. We therefore regard the concentration of narrow absorption near phase~0 as suggestive of orbital or azimuthal structure, rather than as evidence for a strictly phase-locked absorber.

The measured ionization and column density also constrain the characteristic location of the absorbing material. Combining $\xi=L_{\rm ion}/nr^2$ with $N_{\rm H}=n\Delta r$, and requiring $\Delta r\leq r$, gives $r_{\rm max}=L_{\rm ion}/(\xi N_{\rm H})$. Using the measured 0.01--100~keV ionizing flux, $F_{\rm ion}=1.0081\times10^{-9}~{\rm erg\,s^{-1}\,cm^{-2}}$, for which
\begin{equation}
L_{\rm ion}=1.21\times10^{37}\left(\frac{d}{10~{\rm kpc}}\right)^2{\rm erg\,s^{-1}},
\end{equation}
we obtain $r_{\rm max}=0.4^{+4.9}_{-0.3}\times10^{8}\; (d_{10})^2$~cm for Part~2, $7^{+33}_{-5}\times10^{9}(d_{10})^2$~cm for the Narrow spectrum, and $8^{+16}_{-6}\times10^{9}(d_{10})^2$~cm for Phase~0 (where $d_{10}$ represents the distance in units of $10~\rm kpc$). We caution that the particularly small radius inferred for Part~2 is sensitive to the strong degeneracy between the ionization parameter and column density discussed in Section \ref{subsec:time-resolved}. The alternative, lower-ionization and lower-column solution, which provides a virtually identical fit statistic, yields $r_{\max}\simeq4.7\times10^9(d/10\;{\rm kpc})^2~{\rm cm}$, approximately two orders of magnitude larger than the nominal estimate and comparable to the radii inferred from the shorter integrations. Although this alternative radius remains below the characteristic thermal-launching regime for the assumed parameters, the difference illustrates the substantial model dependence of the inferred absorber location.

These values can be compared with the Compton radius, $R_{\rm IC}=GM_{\rm BH}\mu m_p/(k_{\rm B}T_{\rm IC})$, which sets the characteristic scale for thermally driven outflows. Thermal winds can typically be launched at radii $\gtrsim0.1R_{\rm IC}$, although the precise threshold depends on the illuminating spectrum, luminosity, and contribution of radiation pressure (\citealt{1983ApJ...271...70B,2018MNRAS.473..838D}). For $M_{\rm BH}=5\pm1\,M_\odot$ (\citealt{2021RAA....21..214S}) and a Compton temperature of $T_{\rm IC}=9.3\times10^6\;\rm K$, we estimate $R_{\rm IC}=(5\pm1)\times10^{11}$~cm. At an assumed distance of 10~kpc, the inferred upper limits on the absorber radius correspond to approximately $8\times10^{-5}\;R_{\rm IC}$, $0.014\;R_{\rm IC}$, and $0.016\;R_{\rm IC}$ for Part~2, Narrow, and Phase~0, respectively. Although the nominal radius estimates remain below the characteristic thermal-launching regime, their interpretation is limited by the uncertainties in the absorber properties, particularly the degeneracy between ionization and column density in Part~2. These radius constraints, however, remain sensitive to the uncertain distance, absorber geometry, and photoionization parameters.

Several mechanisms may contribute to launching the outflow detected in 4U~1957+115. Thermal driving is one possible mechanism, since sustained X-ray irradiation can heat the disk atmosphere and launch an outflow at sufficiently large radii (\citealt{1983ApJ...271...70B,2018MNRAS.473..838D}). However, the efficiency of this process depends on the irradiating spectrum, luminosity, and location of the absorbing material. The relatively low velocities measured in the Narrow and Phase~0 spectra are also broadly consistent with predictions from thermal-wind simulations (\citealt{2017ApJ...836...42H}), although such velocities are not unique to thermally driven outflows. However, the inferred upper limits on absorber radius remain below the characteristic thermal-launching regime, even for these shorter intervals. Radiation pressure may contribute to the acceleration, although the high ionization of the absorbing gas makes UV line driving unlikely to be efficient, while the importance of electron-scattering radiation pressure depends on the uncertain Eddington ratio. Alternatively, magnetic pressure or magnetocentrifugal forces could accelerate material from the relatively small radii implied by our fits, as proposed for other BH XRB winds (\citealt{2006Natur.441..953M,2017NatAs...1E..62F}). The rapid appearance of absorption despite the nearly constant continuum does not, by itself, distinguish between these mechanisms, since a persistent outflow could exhibit substantial variability in its density, ionization structure, or line-of-sight covering fraction. Thus, although the nominal radius estimates present a challenge for purely thermal driving, the uncertain geometry and physical structure of the absorber prevent us from establishing a unique launching mechanism.

An additional possibility is that the absorbing material does not constitute a freely escaping wind, but instead represents a bound disk atmosphere or a partially accelerated, potentially failed outflow. Under the adopted single-zone radius estimates, the measured line-of-sight velocities of $120-420\;\rm km/s$ are substantially smaller than the corresponding local escape velocities. A similar discrepancy between compact inferred absorber locations and relatively slow outflows has been identified in other BH XRBs, including recent XRISM observations of 4U~1630--472, where some absorption components have been interpreted as failed winds (\citealt{2025ApJ...988L..28M,2026arXiv260908731K}). However, the observed blueshifts measure only the line-of-sight velocity, and the absorbing material may still be accelerating or moving along trajectories substantially inclined to our viewing direction. Furthermore, the radius estimates depend on the assumed geometry and thickness of the absorbing layer, which may not adequately describe an extended or stratified wind. We therefore regard a failed wind as a plausible interpretation, rather than a necessary consequence of the measured velocities.

The inferred wind properties also allow us to estimate its mass outflow rate and kinetic power. These estimates assume a steady, radially outflowing wind; if some of the absorbing material instead forms a bound or failed outflow, the inferred mass flux would not necessarily represent net mass loss from the system. For a photoionized outflow, the mass-loss rate can be written as 
\begin{equation}
    \dot{M}_{\rm wind}
    =
    4\pi C_{\Omega} C_{\rm V}\mu m_{\rm p}
    \frac{L_{\rm ion}}{\xi}v_{\rm out},
\end{equation}
where $C_{\Omega}=\Omega/4\pi$ is the global covering factor and $C_{\rm V}$ is the volume filling factor. The corresponding kinetic luminosity is 
\begin{equation}
    L_{\rm kin}
    =
    \frac{1}{2}\dot{M}_{\rm wind}v_{\rm out}^{2}.
\end{equation}
Using the posterior distributions of $\xi$ and $v_{\rm out}$, we infer $L_{\rm kin}/L_{\rm ion} =1.0^{+7.5}_{-0.8}\times10^{-6} C_{\Omega}C_{\rm V}$ for Part~2, $4.1^{+28.7}_{-3.3}\times10^{-7} C_{\Omega}C_{\rm V}$ for the Narrow interval, and $1.1^{+23.7}_{-1.0}\times10^{-8} C_{\Omega}C_{\rm V}$ for Phase~0. Thus, despite the substantial column densities inferred for the absorber, the relatively low outflow velocities imply that the wind carries only a small fraction of the radiative luminosity in kinetic form. The corresponding absolute kinetic luminosity therefore inherits the same $d^2$ dependence as $L_{\rm ion}$. If we further assume that $L_{\rm bol}\simeq L_{\rm ion}$, the ratio of the wind mass-loss rate to the accretion rate becomes independent of distance:
\begin{equation}
    \frac{\dot{M}_{\rm wind}}{\dot{M}_{\rm acc}}
    =
    \frac{\dot{M}_{\rm wind}}{L_{\rm ion}}\eta c^2.
\end{equation}
These estimates also inherit the uncertainties associated with the photoionization modeling, particularly the strong correlation between ionization and column density in Part~2. Because the mass outflow rate scales inversely with $\xi$, alternative solutions with comparable fit statistics may yield substantially different wind energetics.

For $\eta=0.1$, this gives $\dot{M}_{\rm wind}/\dot{M}_{\rm acc} =0.10^{+0.22}_{-0.06}C_{\Omega}C_V$, $0.41^{+0.69}_{-0.25}C_{\Omega}C_V$, and $0.22^{+0.27}_{-0.17}C_{\Omega}C_V$ for Part~2, Narrow, and Phase~0, respectively. 
Since neither the global covering factor $C_{\Omega}$ nor the volume filling factor $C_V$ can be independently constrained by the present observations, these estimates remain strongly dependent on the assumed wind geometry. Although the inferred clumpiness may suggest a small volume filling factor, it does not necessarily imply a small global covering factor. For the nominal spectral solutions, the inferred mass-loss rates remain below the accretion rate even when $C_\Omega C_V=1$. However, the alternative Part~2 solution permits a substantially larger mass-loss rate, while the unknown wind geometry introduces additional uncertainty. We therefore cannot robustly determine the fraction of accreting material removed by the outflow.

This may be particularly interesting in the context of the persistent nature of 4U~1957+115. Within the irradiated disk-instability model, the source is expected to remain persistent if its mass-transfer rate is sufficiently high to keep the outer disk hot and ionized, preventing the thermal-viscous instability that drives transient outbursts (\citealt{2012MNRAS.424.1991C}). Strong wind mass loss can instead deplete the disk and shorten X-ray binary outbursts, whereas scattering of the central X-ray emission in an extended wind can enhance irradiation of the outer disk and promote its thermal stability (\citealt{2019A&A...632A..40D,2020MNRAS.495.3666T}). The influence of the wind on the persistent accretion state of 4U~1957+115 therefore depends on the balance between mass loss and the additional irradiation provided by scattering in the outflow. However, neither the global wind geometry nor its scattering efficiency is constrained by the present observations, preventing us from determining whether the wind plays a significant role in maintaining the source's persistent soft state.

Taken together, the XRISM observation reveals a disk wind in 4U~1957+115 that is highly structured in both time and velocity space. The absorber changes substantially over the course of the observation, while the shortest time- and phase-resolved spectra reveal considerably narrower Fe~XXV and Fe~XXVI features than those inferred from longer integrations. In particular, the turbulent velocity decreases from $v_{\rm turb}=350^{+290}_{-170}~{\rm km/s}$ in Part~2 to $120^{+120}_{-80}~{\rm km/s}$ during the Narrow interval and to $v_{\rm turb}\lesssim70~{\rm km/s}$ around phase~0. This evolution suggests that line broadening measured in time-averaged spectra may contain a contribution from unresolved changes in the velocity and structure of the absorbing material, rather than tracing intrinsic turbulence alone. The spectra also contain hints of asymmetric absorption and P-Cygni-like profiles, although neither is statistically significant when tested with additional spectral components. The combination of the spectral resolution of XRISM/Resolve with time-resolved spectroscopy therefore provides a powerful means of investigating the relative contributions of intrinsic line broadening and macroscopic wind variability. Future observations that sample multiple binary orbits with higher signal-to-noise ratios will be particularly valuable for determining whether the changes observed here arise from accelerating clumps, persistent azimuthal structure, or orbital modulation of the wind, and for searching for more subtle signatures of wind geometry and acceleration.

\vspace{-0.5cm}
\begin{acknowledgments}
P.D. thanks Yerong Xu and Ciro Pinto for their help with the SPEX-to-XSPEC package. E.B. is supported by The Israel Science Foundation (Grant No. 2617/25). V.B. is supported by the National Science Foundation Graduate Research Fellowship Program (Grant No. 2141064).

The authors declare the use of generative AI in the research and writing process. According to the GAIDeT taxonomy (2026), the following tasks were delegated to GAI tools under full human supervision: Code optimization, Process automation, Visualization, Proofreading and editing the text for clarity. 

The GAI tool used was ChatGPT GPT-6 Astra.
Responsibility for the final manuscript lies entirely with the authors.
GAI tools are not listed as authors and do not bear responsibility for the final outcomes.
Declaration submitted by Paul Draghis
\end{acknowledgments}

\bibliography{sample7}{}

\begin{thebibliography}{}
\expandafter\ifx\csname natexlab\endcsname\relax\def\natexlab#1{#1}\fi
\providecommand{\url}[1]{\href{#1}{#1}}
\providecommand{\dodoi}[1]{doi:~\href{http://doi.org/#1}{\nolinkurl{#1}}}
\providecommand{\doeprint}[1]{\href{http://ascl.net/#1}{\nolinkurl{http://ascl.net/#1}}}
\providecommand{\doarXiv}[1]{\href{https://arxiv.org/abs/#1}{\nolinkurl{https://arxiv.org/abs/#1}}}

\bibitem[{H. {Akaike}(1974){Akaike}}]{1974ITAC...19..716A}
{Akaike}, H. 1974, \bibinfo{title}{{A New Look at the Statistical Model Identification},} IEEE Transactions on Automatic Control, 19, 716

\bibitem[{P.~J. {Armitage} \& M. {Livio}(1996){Armitage} \& {Livio}}]{1996ApJ...470.1024A}
{Armitage}, P.~J., \& {Livio}, M. 1996, \bibinfo{title}{{Accretion Disks in Interacting Binaries: Simulations of the Stream-Disk Impact},} \apj, 470, 1024, \dodoi{10.1086/177928}

\bibitem[{E. {Barillier} {et~al.}(2023){Barillier}, {Grinberg}, {Horn}, {Nowak}, {Remillard}, {Steiner}, {Walton}, \& {Wilms}}]{2023ApJ...944..165B}
{Barillier}, E., {Grinberg}, V., {Horn}, D., {et~al.} 2023, \bibinfo{title}{{NICER/NuSTAR Characterization of 4U 1957+11: A Near Maximally Spinning Black Hole Potentially in the Mass Gap},} \apj, 944, 165, \dodoi{10.3847/1538-4357/acaeaf}

\bibitem[{M.~C. {Begelman} {et~al.}(1983){Begelman}, {McKee}, \& {Shields}}]{1983ApJ...271...70B}
{Begelman}, M.~C., {McKee}, C.~F., \& {Shields}, G.~A. 1983, \bibinfo{title}{{Compton heated winds and coronae above accretion disks. I. Dynamics},} \apj, 271, 70, \dodoi{10.1086/161178}

\bibitem[{M. {Coriat} {et~al.}(2012){Coriat}, {Fender}, \& {Dubus}}]{2012MNRAS.424.1991C}
{Coriat}, M., {Fender}, R.~P., \& {Dubus}, G. 2012, \bibinfo{title}{{Revisiting a fundamental test of the disc instability model for X-ray binaries},} \mnras, 424, 1991, \dodoi{10.1111/j.1365-2966.2012.21339.x}

\bibitem[{T. {Dauser} {et~al.}(2014){Dauser}, {Garcia}, {Parker}, {Fabian}, \& {Wilms}}]{2014MNRAS.444L.100D}
{Dauser}, T., {Garcia}, J., {Parker}, M.~L., {Fabian}, A.~C., \& {Wilms}, J. 2014, \bibinfo{title}{{The role of the reflection fraction in constraining black hole spin.},} \mnras, 444, L100, \dodoi{10.1093/mnrasl/slu125}

\bibitem[{M. {D{\'\i}az Trigo} \& L. {Boirin}(2016){D{\'\i}az Trigo} \& {Boirin}}]{2016AN....337..368D}
{D{\'\i}az Trigo}, M., \& {Boirin}, L. 2016, \bibinfo{title}{{Accretion disc atmospheres and winds in low-mass X-ray binaries},} Astronomische Nachrichten, 337, 368, \dodoi{10.1002/asna.201612315}

\bibitem[{C. {Done} {et~al.}(2018){Done}, {Tomaru}, \& {Takahashi}}]{2018MNRAS.473..838D}
{Done}, C., {Tomaru}, R., \& {Takahashi}, T. 2018, \bibinfo{title}{{Thermal winds in stellar mass black hole and neutron star binary systems},} \mnras, 473, 838, \dodoi{10.1093/mnras/stx2400}

\bibitem[{P.~A. {Draghis} {et~al.}(2023){Draghis}, {Miller}, {Zoghbi}, {Reynolds}, {Costantini}, {Gallo}, \& {Tomsick}}]{2023ApJ...946...19D}
{Draghis}, P.~A., {Miller}, J.~M., {Zoghbi}, A., {et~al.} 2023, \bibinfo{title}{{A Systematic View of Ten New Black Hole Spins},} \apj, 946, 19, \dodoi{10.3847/1538-4357/acafe7}

\bibitem[{G. {Dubus} {et~al.}(2019){Dubus}, {Done}, {Tetarenko}, \& {Hameury}}]{2019A&A...632A..40D}
{Dubus}, G., {Done}, C., {Tetarenko}, B.~E., \& {Hameury}, J.-M. 2019, \bibinfo{title}{{The impact of thermal winds on the outburst lightcurves of black hole X-ray binaries},} \aap, 632, A40, \dodoi{10.1051/0004-6361/201936333}

\bibitem[{K. {Fukumura} {et~al.}(2017){Fukumura}, {Kazanas}, {Shrader}, {Behar}, {Tombesi}, \& {Contopoulos}}]{2017NatAs...1E..62F}
{Fukumura}, K., {Kazanas}, D., {Shrader}, C., {et~al.} 2017, \bibinfo{title}{{Magnetic origin of black hole winds across the mass scale},} Nature Astronomy, 1, 0062, \dodoi{10.1038/s41550-017-0062}

\bibitem[{J. {Garc{\'\i}a} {et~al.}(2014){Garc{\'\i}a}, {Dauser}, {Lohfink}, {Kallman}, {Steiner}, {McClintock}, {Brenneman}, {Wilms}, {Eikmann}, {Reynolds}, \& {Tombesi}}]{2014ApJ...782...76G}
{Garc{\'\i}a}, J., {Dauser}, T., {Lohfink}, A., {et~al.} 2014, \bibinfo{title}{{Improved Reflection Models of Black Hole Accretion Disks: Treating the Angular Distribution of X-Rays},} \apj, 782, 76, \dodoi{10.1088/0004-637X/782/2/76}

\bibitem[{P. {Hakala} {et~al.}(2014){Hakala}, {Muhli}, \& {Charles}}]{2014MNRAS.444.3802H}
{Hakala}, P., {Muhli}, P., \& {Charles}, P. 2014, \bibinfo{title}{{Simultaneous optical and near-IR photometry of 4U1957+115 - a missing secondary star},} \mnras, 444, 3802, \dodoi{10.1093/mnras/stu1687}

\bibitem[{N. {Higginbottom} {et~al.}(2017){Higginbottom}, {Proga}, {Knigge}, \& {Long}}]{2017ApJ...836...42H}
{Higginbottom}, N., {Proga}, D., {Knigge}, C., \& {Long}, K.~S. 2017, \bibinfo{title}{{Thermal Disk Winds in X-Ray Binaries: Realistic Heating and Cooling Rates Give Rise to Slow, but Massive, Outflows},} \apj, 836, 42, \dodoi{10.3847/1538-4357/836/1/42}

\bibitem[{Y. {Ishisaki} {et~al.}(2025){Ishisaki}, {Kelley}, {Awaki}, {Balleza}, {Barnstable}, {Bialas}, {Boissay-Malaquin}, {Brown}, {Canavan}, {Cumbee}, {Carnahan}, {Chiao}, {Comber}, {Costantini}, {den Herder}, {Dercksen}, {de Vries}, {DiPirro}, {Eckart}, {Ezoe}, {Ferrigno}, {Fujimoto}, {Gorter}, {Graham}, {Grim}, {Hartz}, {Hayakawa}, {Hayashi}, {Hell}, {Hoshino}, {Ichinohe}, {Ishida}, {Ishikawa}, {James}, {Kenyon}, {Kilbourne}, {Kimball}, {Kitamoto}, {Leutenegger}, {Maeda}, {McCammon}, {Miko}, {Mizumoto}, {Noda}, {Okajima}, {Okamoto}, {Paltani}, {Porter}, {Sato}, {Sato}, {Sawada}, {Shinozaki}, {Shipman}, {Shirron}, {Sneiderman}, {Soong}, {Szymkiewicz}, {Szymkowiak}, {Takei}, {Tamura}, {Tsujimoto}, {Uchida}, {Wasserzug}, {Witthoeft}, {Wolfs}, {Yamada}, \& {Yasuda}}]{2025JATIS..11d2023I}
{Ishisaki}, Y., {Kelley}, R.~L., {Awaki}, H., {et~al.} 2025, \bibinfo{title}{{Resolve instrument onboard XRISM: design, integration, and instrument test results},} Journal of Astronomical Telescopes, Instruments, and Systems, 11, 042023, \dodoi{10.1117/1.JATIS.11.4.042023}

\bibitem[{J.~S. {Kaastra} \& J.~A.~M. {Bleeker}(2016){Kaastra} \& {Bleeker}}]{2016A&A...587A.151K}
{Kaastra}, J.~S., \& {Bleeker}, J.~A.~M. 2016, \bibinfo{title}{{Optimal binning of X-ray spectra and response matrix design},} \aap, 587, A151, \dodoi{10.1051/0004-6361/201527395}

\bibitem[{N. {Keshet} {et~al.}(2026){Keshet}, {Behar}, {Miller}, \& {Neilsen}}]{2026arXiv260908731K}
{Keshet}, N., {Behar}, E., {Miller}, J.~M., \& {Neilsen}, J. 2026, \bibinfo{title}{{Stratified Wind Geometries of GRS 1915+105: Density Profiles and Radii},} arXiv e-prints, arXiv:2609.08731, \dodoi{10.48550/arXiv.2609.08731}

\bibitem[{H. {Liu} {et~al.}(2022){Liu}, {Fu}, {Bambi}, {Jiang}, {Parker}, {Ji}, {Kong}, {Zhang}, {Zhang}, \& {Zhang}}]{2022ApJ...933..122L}
{Liu}, H., {Fu}, Y., {Bambi}, C., {et~al.} 2022, \bibinfo{title}{{The Disk Wind in GRS 1915+105 as Seen by Insight-Hard X-Ray Modulation Telescope},} \apj, 933, 122, \dodoi{10.3847/1538-4357/ac74b1}

\bibitem[{L. {Marra} {et~al.}(2024){Marra}, {Brigitte}, {Rodriguez Cavero}, {Chun}, {Steiner}, {Dov{\v{c}}iak}, {Nowak}, {Bianchi}, {Capitanio}, {Ingram}, {Matt}, {Muleri}, {Podgorn{\'y}}, {Poutanen}, {Svoboda}, {Taverna}, {Ursini}, {Veledina}, {De Rosa}, {Garc{\'\i}a}, {Lutovinov}, {Mereminskiy}, {Farinelli}, {Gunji}, {Kaaret}, {Kallman}, {Krawczynski}, {Kan}, {Hu}, {Marinucci}, {Mastroserio}, {Mikus̆incov{\'a}}, {Parra}, {Petrucci}, {Ratheesh}, {Soffitta}, {Tombesi}, {Zane}, {Agudo}, {Antonelli}, {Bachetti}, {Baldini}, {Baumgartner}, {Bellazzini}, {Bongiorno}, {Bonino}, {Brez}, {Bucciantini}, {Castellano}, {Cavazzuti}, {Chen}, {Ciprini}, {Costa}, {Del Monte}, {Di Gesu}, {Di Lalla}, {Di Marco}, {Donnarumma}, {Doroshenko}, {Ehlert}, {Enoto}, {Evangelista}, {Fabiani}, {Ferrazzoli}, {Hayashida}, {Heyl}, {Iwakiri}, {Jorstad}, {Karas}, {Kislat}, {Kitaguchi}, {Kolodziejczak}, {La Monaca}, {Latronico}, {Liodakis}, {Maldera}, {Manfreda}, {Marin}, {Marscher}, {Marshall}, {Massaro}, {Mitsuishi}, {Mizuno}, {Negro},
  {Ng}, {O'Dell}, {Omodei}, {Oppedisano}, {Papitto}, {Pavlov}, {Peirson}, {Perri}, {Pesce-Rollins}, {Pilia}, {Possenti}, {Puccetti}, {Ramsey}, {Rankin}, {Roberts}, {Romani}, {Sgr{\`o}}, {Slane}, {Spandre}, {Swartz}, {Tamagawa}, {Tavecchio}, {Tawara}, {Tennant}, {Thomas}, {Trois}, {Tsygankov}, {Turolla}, {Vink}, {Weisskopf}, {Wu}, \& {Xie}}]{2024A&A...684A..95M}
{Marra}, L., {Brigitte}, M., {Rodriguez Cavero}, N., {et~al.} 2024, \bibinfo{title}{{IXPE observation confirms a high spin in the accreting black hole 4U 1957+115},} \aap, 684, A95, \dodoi{10.1051/0004-6361/202348277}

\bibitem[{P.~A. {Mason} {et~al.}(2012){Mason}, {Robinson}, {Bayless}, \& {Hakala}}]{2012AJ....144..108M}
{Mason}, P.~A., {Robinson}, E.~L., {Bayless}, A.~J., \& {Hakala}, P.~J. 2012, \bibinfo{title}{{Long-term Optical Observations of Two LMXBS: UW CrB (=MS 1603+260) and V1408 Aql (=4U 1957+115)},} \aj, 144, 108, \dodoi{10.1088/0004-6256/144/4/108}

\bibitem[{M. {Mehdipour} {et~al.}(2016){Mehdipour}, {Kaastra}, \& {Kallman}}]{2016A&A...596A..65M}
{Mehdipour}, M., {Kaastra}, J.~S., \& {Kallman}, T. 2016, \bibinfo{title}{{Systematic comparison of photoionised plasma codes with application to spectroscopic studies of AGN in X-rays},} \aap, 596, A65, \dodoi{10.1051/0004-6361/201628721}

\bibitem[{J.~M. {Miller} {et~al.}(2006{\natexlab{a}}){Miller}, {Raymond}, {Fabian}, {Steeghs}, {Homan}, {Reynolds}, {van der Klis}, \& {Wijnands}}]{2006Natur.441..953M}
{Miller}, J.~M., {Raymond}, J., {Fabian}, A.~C., {et~al.} 2006{\natexlab{a}}, \bibinfo{title}{{The magnetic nature of disk accretion onto black holes},} \nat, 441, 953, \dodoi{10.1038/nature04912}

\bibitem[{J.~M. {Miller} {et~al.}(2006{\natexlab{b}}){Miller}, {Raymond}, {Homan}, {Fabian}, {Steeghs}, {Wijnands}, {Rupen}, {Charles}, {van der Klis}, \& {Lewin}}]{2006ApJ...646..394M}
{Miller}, J.~M., {Raymond}, J., {Homan}, J., {et~al.} 2006{\natexlab{b}}, \bibinfo{title}{{Simultaneous Chandra and RXTE Spectroscopy of the Microquasar H1743-322: Clues to Disk Wind and Jet Formation from a Variable Ionized Outflow},} \apj, 646, 394, \dodoi{10.1086/504673}

\bibitem[{J.~M. {Miller} {et~al.}(2015){Miller}, {Kaastra}, {Miller}, {Reynolds}, {Brown}, {Cenko}, {Drake}, {Gezari}, {Guillochon}, {Gultekin}, {Irwin}, {Levan}, {Maitra}, {Maksym}, {Mushotzky}, {O'Brien}, {Paerels}, {de Plaa}, {Ramirez-Ruiz}, {Strohmayer}, \& {Tanvir}}]{2015Natur.526..542M}
{Miller}, J.~M., {Kaastra}, J.~S., {Miller}, M.~C., {et~al.} 2015, \bibinfo{title}{{Flows of X-ray gas reveal the disruption of a star by a massive black hole},} \nat, 526, 542, \dodoi{10.1038/nature15708}

\bibitem[{J.~M. {Miller} {et~al.}(2016){Miller}, {Raymond}, {Fabian}, {Gallo}, {Kaastra}, {Kallman}, {King}, {Proga}, {Reynolds}, \& {Zoghbi}}]{2016ApJ...821L...9M}
{Miller}, J.~M., {Raymond}, J., {Fabian}, A.~C., {et~al.} 2016, \bibinfo{title}{{The Accretion Disk Wind in the Black Hole GRS 1915+105},} \apjl, 821, L9, \dodoi{10.3847/2041-8205/821/1/L9}

\bibitem[{J.~M. {Miller} {et~al.}(2025){Miller}, {Mizumoto}, {Shidatsu}, {Ballhausen}, {Behar}, {D{\'\i}az Trigo}, {Done}, {Dotani}, {Garc{\'\i}a}, {Kallman}, {Kobayashi}, {Kubota}, {Smith}, {Takahashi}, {Tashiro}, {Ueda}, {Vink}, {Yamada}, {Watanabe}, {Iizuka}, {Terada}, {Baluta}, {Kanemaru}, {Ogawa}, {Yoshida}, \& {Hayashi}}]{2025ApJ...988L..28M}
{Miller}, J.~M., {Mizumoto}, M., {Shidatsu}, M., {et~al.} 2025, \bibinfo{title}{{XRISM Spectroscopy of the Stellar-mass Black Hole 4U 1630-472 in Outburst},} \apjl, 988, L28, \dodoi{10.3847/2041-8213/ade25c}

\bibitem[{J. {Neilsen} \& J.~C. {Lee}(2009){Neilsen} \& {Lee}}]{2009Natur.458..481N}
{Neilsen}, J., \& {Lee}, J.~C. 2009, \bibinfo{title}{{Accretion disk winds as the jet suppression mechanism in the microquasar GRS 1915+105},} \nat, 458, 481, \dodoi{10.1038/nature07680}

\bibitem[{J. {Neilsen} {et~al.}(2012){Neilsen}, {Petschek}, \& {Lee}}]{2012MNRAS.421..502N}
{Neilsen}, J., {Petschek}, A.~J., \& {Lee}, J.~C. 2012, \bibinfo{title}{{Accretion disc wind variability in the states of the microquasar GRS 1915+105},} \mnras, 421, 502, \dodoi{10.1111/j.1365-2966.2011.20329.x}

\bibitem[{J. {Neilsen} {et~al.}(2018){Neilsen}, {Cackett}, {Remillard}, {Homan}, {Steiner}, {Gendreau}, {Arzoumanian}, {Prigozhin}, {LaMarr}, {Doty}, {Eikenberry}, {Tombesi}, {Ludlam}, {Kara}, {Altamirano}, \& {Fabian}}]{2018ApJ...860L..19N}
{Neilsen}, J., {Cackett}, E.~M., {Remillard}, R.~A., {et~al.} 2018, \bibinfo{title}{{A Persistent Disk Wind in GRS 1915+105 with NICER},} \apjl, 860, L19, \dodoi{10.3847/2041-8213/aacc6c}

\bibitem[{M.~A. {Nowak} {et~al.}(2008){Nowak}, {Juett}, {Homan}, {Yao}, {Wilms}, {Schulz}, \& {Canizares}}]{2008ApJ...689.1199N}
{Nowak}, M.~A., {Juett}, A., {Homan}, J., {et~al.} 2008, \bibinfo{title}{{Disk-dominated States of 4U 1957+11: Chandra, XMM-Newton, and RXTE Observations of Ostensibly the Most Rapidly Spinning Galactic Black Hole},} \apj, 689, 1199, \dodoi{10.1086/592227}

\bibitem[{M.~L. {Parker} {et~al.}(2019){Parker}, {Longinotti}, {Schartel}, {Grupe}, {Komossa}, {Kriss}, {Fabian}, {Gallo}, {Harrison}, {Jiang}, {Kara}, {Krongold}, {Matzeu}, {Pinto}, \& {Santos-Lle{\'o}}}]{2019MNRAS.490..683P}
{Parker}, M.~L., {Longinotti}, A.~L., {Schartel}, N., {et~al.} 2019, \bibinfo{title}{{The nuclear environment of the NLS1 Mrk 335: Obscuration of the X-ray line emission by a variable outflow},} \mnras, 490, 683, \dodoi{10.1093/mnras/stz2566}

\bibitem[{G. {Ponti} {et~al.}(2012){Ponti}, {Fender}, {Begelman}, {Dunn}, {Neilsen}, \& {Coriat}}]{2012MNRAS.422L..11P}
{Ponti}, G., {Fender}, R.~P., {Begelman}, M.~C., {et~al.} 2012, \bibinfo{title}{{Ubiquitous equatorial accretion disc winds in black hole soft states},} \mnras, 422, L11.
\newblock \doarXiv{1201.4172}

\bibitem[{G. {Schwarz}(1978){Schwarz}}]{1978AnSta...6..461S}
{Schwarz}, G. 1978, \bibinfo{title}{{Estimating the Dimension of a Model},} Annals of Statistics, 6, 461

\bibitem[{P. {Sharma} {et~al.}(2021){Sharma}, {Sharma}, {Jain}, {Dewangan}, \& {Dutta}}]{2021RAA....21..214S}
{Sharma}, P., {Sharma}, R., {Jain}, C., {Dewangan}, G.~C., \& {Dutta}, A. 2021, \bibinfo{title}{{Broad-band spectral study of LMXB black hole candidate 4U 1957+11 with NuSTAR},} Research in Astronomy and Astrophysics, 21, 214, \dodoi{10.1088/1674-4527/21/9/214}

\bibitem[{M. {Tashiro} {et~al.}(2021){Tashiro}, {Maejima}, {Toda}, {Kelley}, {Reichenthal}, {Hartz}, {Petre}, {Williams}, {Guainazzi}, {Costantini}, {Fujimoto}, {Hayashida}, {Henegar-Leon}, {Holland}, {Ishisaki}, {Kilbourne}, {Loewenstein}, {Matsushita}, {Mori}, {Okajima}, {Porter}, {Sneiderman}, {Takei}, {Terada}, {Tomida}, {Yamaguchi}, {Watanabe}, {Akamatsu}, {Arai}, {Audard}, {Awaki}, {Babyk}, {Bamba}, {Bando}, {Behar}, {Bialas}, {Boissay-Malaquin}, {Brenneman}, {Brown}, {Canavan}, {Chiao}, {Comber}, {Corrales}, {Cumbee}, {de Vries}, {den Herder}, {Dercksen}, {Diaz-Trigo}, {DiPirro}, {Done}, {Dotani}, {Ebisawa}, {Eckart}, {Eckert}, {Eguchi}, {Enoto}, {Ezoe}, {Ferrigno}, {Fujita}, {Fukazawa}, {Furuzawa}, {Gallo}, {Gorter}, {Grim}, {Gu}, {Hagino}, {Hamaguchi}, {Hatsukade}, {Hawthorn}, {Hayashi}, {Hell}, {Hiraga}, {Hodges-Kluck}, {Horiuchi}, {Hornschemeier}, {Hoshino}, {Ichinohe}, {Iga}, {Iizuka}, {Ishida}, {Ishihama}, {Ishikawa}, {Ishimura}, {Jaffe}, {Kaastra}, {Kallman}, {Kara}, {Katsuda}, {Kenyon},
  {Kimball}, {Kitaguchi}, {Kitamoto}, {Kobayashi}, {Kobayashi}, {Kohmura}, {Kubota}, {Leutenegger}, {Li}, {Lockard}, {Maeda}, {Markevitch}, {Martz}, {Matsumoto}, {Matsuzaki}, {McCammon}, {McLaughlin}, {McNamara}, {Miko}, {Miller}, {Miller}, {Minesugi}, {Mitani}, {Mitsuishi}, {Mizumoto}, {Mizuno}, {Mukai}, {Murakami}, {Mushotzky}, {Nakajima}, {Nakamura}, {Nakazawa}, {Natsukari}, {Nigo}, {Nishioka}, {Nobukawa}, {Nobukawa}, {Noda}, {Odaka}, {Ogawa}, {Ohashi}, {Ohno}, {Ohta}, {Okamoto}, {Ota}, {Ozaki}, {Paltani}, {Plucinsky}, {Pottschmidt}, {Sampson}, {Sasaki}, {Sato}, {Sato}, {Sato}, {Sawada}, {Seta}, {Shibano}, {Shida}, {Shidatsu}, {Shigeto}, {Shinozaki}, {Shirron}, {Simionescu}, {Smith}, {Someya}, {Soong}, {Sugawara}, {Sugawara}, {Szymkowiak}, {Takahashi}, {Takeshima}, {Tamagawa}, {Tamura}, {Tanaka}, {Tanimoto}, {Terashima}, {Tsuboi}, {Tsujimoto}, {Tsunemi}, {Tsuru}, {Uchida}, {Uchida}, {Uchiyama}, {Ueda}, {Uno}, {Vink}, {Watanabe}, {Witthoeft}, {Wolfs}, {Yamada}, {Yamaoka}, {Yamasaki}, {Yamauchi}, {Yamauchi},
  {Yanagase}, {Yaqoob}, {Yasuda}, {Yoshida}, {Yoshioka}, \& {Zhuravleva}}]{2021SPIE11444E..22T}
{Tashiro}, M., {Maejima}, H., {Toda}, K., {et~al.} 2021, in Society of Photo-Optical Instrumentation Engineers (SPIE) Conference Series, Vol. 11444, Society of Photo-Optical Instrumentation Engineers (SPIE) Conference Series, ed. J.-W.~A. {den Herder}, S.~{Nikzad}, \& K.~{Nakazawa}, 1144422, \dodoi{10.1117/12.2565812}

\bibitem[{B.~E. {Tetarenko} {et~al.}(2020){Tetarenko}, {Dubus}, {Marcel}, {Done}, \& {Clavel}}]{2020MNRAS.495.3666T}
{Tetarenko}, B.~E., {Dubus}, G., {Marcel}, G., {Done}, C., \& {Clavel}, M. 2020, \bibinfo{title}{{Thermally driven disc winds as a mechanism for X-ray irradiation heating in black hole X-ray binaries: the case study of GX339-4},} \mnras, 495, 3666, \dodoi{10.1093/mnras/staa1367}

\bibitem[{R. {Tomaru} {et~al.}(2019){Tomaru}, {Done}, {Ohsuga}, {Nomura}, \& {Takahashi}}]{2019MNRAS.490.3098T}
{Tomaru}, R., {Done}, C., {Ohsuga}, K., {Nomura}, M., \& {Takahashi}, T. 2019, \bibinfo{title}{{The thermal-radiative wind in low-mass X-ray binary H1743-322: radiation hydrodynamic simulations},} \mnras, 490, 3098, \dodoi{10.1093/mnras/stz2738}

\bibitem[{R. {Wijnands} {et~al.}(2002){Wijnands}, {Miller}, \& {van der Klis}}]{2002MNRAS.331...60W}
{Wijnands}, R., {Miller}, J.~M., \& {van der Klis}, M. 2002, \bibinfo{title}{{4U 1957+11: a persistent low-mass X-ray binary and black hole candidate in the high state?},} \mnras, 331, 60, \dodoi{10.1046/j.1365-8711.2002.05137.x}

\bibitem[{Y. {Xu} {et~al.}(2024){Xu}, {Pinto}, {Rogantini}, {Barret}, {Bianchi}, {Guainazzi}, {Ebrero}, {Alston}, {Kara}, \& {Cusumano}}]{2024A&A...687A.179X}
{Xu}, Y., {Pinto}, C., {Rogantini}, D., {et~al.} 2024, \bibinfo{title}{{A systematic study of the ultra-fast outflow responses to luminosity variations in active galactic nuclei},} \aap, 687, A179, \dodoi{10.1051/0004-6361/202349001}

\bibitem[{S. {Yamada} {et~al.}(2025){Yamada}, {Hell}, {Costantini}, {Adegoke}, {Brumback}, {Draghis}, {Ebisawa}, {Garcia}, {Hodges-Kluck}, {Kitamoto}, {Kobayashi}, {Kohmura}, {Kubota}, {Miller}, {Mizumoto}, {Mizuno}, {Ninoyu}, {Takahashi}, {Uchida}, {Yamaoka}, \& {Zhang}}]{2025PASJ...77.1210Y}
{Yamada}, S., {Hell}, N., {Costantini}, E., {et~al.} 2025, \bibinfo{title}{{XRISM high-resolution X-ray spectroscopy of Cygnus X-1: Highly ionized iron absorption structures},} \pasj, 77, 1210, \dodoi{10.1093/pasj/psaf104}

\bibitem[{Y. {Yao} {et~al.}(2008){Yao}, {Nowak}, {Wang}, {Schulz}, \& {Canizares}}]{2008ApJ...672L..21Y}
{Yao}, Y., {Nowak}, M.~A., {Wang}, Q.~D., {Schulz}, N.~S., \& {Canizares}, C.~R. 2008, \bibinfo{title}{{Limits on Hot Galactic Halo Gas from X-Ray Absorption Lines},} \apjl, 672, L21, \dodoi{10.1086/526767}

\bibitem[{S.-N. {Zhang} {et~al.}(2012){Zhang}, {Liao}, \& {Yao}}]{2012MNRAS.421.3550Z}
{Zhang}, S.-N., {Liao}, J., \& {Yao}, Y. 2012, \bibinfo{title}{{Measuring the black hole masses in accreting X-ray binaries by detecting the Doppler orbital motion of their accretion disc wind absorption lines},} \mnras, 421, 3550, \dodoi{10.1111/j.1365-2966.2012.20579.x}

\end{thebibliography}
\bibliographystyle{aasjournalv7}

\appendix 

\section{Line search results}\label{sec:line_search_results}
Table \ref{tab:single-line-search} displays all energies returned through our line-search algorithm for a single \texttt{gaussian} feature. We are only displaying the energies for which at least one of either the time averaged, the Part 1, or the Part 2 spectra return at least one significant ($\geq 3\sigma$) detection at either line width. For an in-depth explanation of the algorithm and the interpretation of the findings, see Section \ref{subsec:line_search}.

\begin{deluxetable}{c| c c c c| c c c c| c c c c| l}[h!]
\tablecaption{Single Gaussian line search results\label{tab:single-line-search}}
\tablehead{
\colhead{Energy} & \multicolumn4c{Time averaged} & \multicolumn4c{First 69 ks} & \multicolumn4c{Last 39 ks} & \colhead{Type} \\
\colhead{(keV)} & \multicolumn4c{Local $\sigma$} & \multicolumn4c{Local $\sigma$} & \multicolumn4c{Local $\sigma$} & \colhead{}}
\startdata
2.394 & \cellcolor{vFifty}1.35 & \cellcolor{vTwoHundred}1.52 & \cellcolor{vFiveHundred}1.77 & \cellcolor{vThousand}1.92 & \cellcolor{vFifty}0.00 & \cellcolor{vTwoHundred}0.00 & \cellcolor{vFiveHundred}0.10 & \cellcolor{vThousand}0.65 & \cellcolor{vFifty}1.82 & \cellcolor{vTwoHundred}2.30 & \cellcolor{vFiveHundred}\textbf{3.05} & \cellcolor{vThousand}2.41 & Abs. \\
2.410 & \cellcolor{vFifty}1.42 & \cellcolor{vTwoHundred}1.58 & \cellcolor{vFiveHundred}1.85 & \cellcolor{vThousand}1.91 & \cellcolor{vFifty}2.64 & \cellcolor{vTwoHundred}2.94 & \cellcolor{vFiveHundred}\textbf{3.24} & \cellcolor{vThousand}2.70 & \cellcolor{vFifty}0.95 & \cellcolor{vTwoHundred}0.96 & \cellcolor{vFiveHundred}0.82 & \cellcolor{vThousand}0.10 & Abs. \\
2.538 & \cellcolor{vFifty}2.55 & \cellcolor{vTwoHundred}2.33 & \cellcolor{vFiveHundred}1.60 & \cellcolor{vThousand}1.16 & \cellcolor{vFifty}\textbf{4.09} & \cellcolor{vTwoHundred}\textbf{4.01} & \cellcolor{vFiveHundred}\textbf{3.20} & \cellcolor{vThousand}2.46 & \cellcolor{vFifty}0.81 & \cellcolor{vTwoHundred}0.85 & \cellcolor{vFiveHundred}1.13 & \cellcolor{vThousand}1.02 & Abs. \\
2.646 & \cellcolor{vFifty}2.52 & \cellcolor{vTwoHundred}2.45 & \cellcolor{vFiveHundred}1.87 & \cellcolor{vThousand}1.22 & \cellcolor{vFifty}\textbf{3.27} & \cellcolor{vTwoHundred}\textbf{3.26} & \cellcolor{vFiveHundred}2.88 & \cellcolor{vThousand}2.10 & \cellcolor{vFifty}0.62 & \cellcolor{vTwoHundred}0.24 & \cellcolor{vFiveHundred}0.39 & \cellcolor{vThousand}0.62 & Em.\\
2.780 & \cellcolor{vFifty}1.48 & \cellcolor{vTwoHundred}1.36 & \cellcolor{vFiveHundred}0.89 & \cellcolor{vThousand}0.75 & \cellcolor{vFifty}\textbf{3.29} & \cellcolor{vTwoHundred}\textbf{3.07} & \cellcolor{vFiveHundred}2.20 & \cellcolor{vThousand}1.71 & \cellcolor{vFifty}1.08 & \cellcolor{vTwoHundred}1.06 & \cellcolor{vFiveHundred}1.09 & \cellcolor{vThousand}0.85 & Em.\\
2.868 & \cellcolor{vFifty}1.01 & \cellcolor{vTwoHundred}1.62 & \cellcolor{vFiveHundred}2.28 & \cellcolor{vThousand}2.08 & \cellcolor{vFifty}0.88 & \cellcolor{vTwoHundred}1.51 & \cellcolor{vFiveHundred}2.68 & \cellcolor{vThousand}\textbf{3.01} & \cellcolor{vFifty}0.89 & \cellcolor{vTwoHundred}0.71 & \cellcolor{vFiveHundred}0.14 & \cellcolor{vThousand}0.72 & Em.\\
2.930 & \cellcolor{vFifty}1.63 & \cellcolor{vTwoHundred}1.79 & \cellcolor{vFiveHundred}2.52 & \cellcolor{vThousand}2.75 & \cellcolor{vFifty}2.71 & \cellcolor{vTwoHundred}\textbf{3.01} & \cellcolor{vFiveHundred}\textbf{3.59} & \cellcolor{vThousand}\textbf{3.06} & \cellcolor{vFifty}1.20 & \cellcolor{vTwoHundred}0.98 & \cellcolor{vFiveHundred}0.20 & \cellcolor{vThousand}0.55 & Em.\\
3.626 & \cellcolor{vFifty}\textbf{3.27} & \cellcolor{vTwoHundred}\textbf{3.62} & \cellcolor{vFiveHundred}2.83 & \cellcolor{vThousand}2.11 & \cellcolor{vFifty}2.80 & \cellcolor{vTwoHundred}\textbf{3.26} & \cellcolor{vFiveHundred}2.50 & \cellcolor{vThousand}2.01 & \cellcolor{vFifty}1.70 & \cellcolor{vTwoHundred}1.35 & \cellcolor{vFiveHundred}1.10 & \cellcolor{vThousand}0.76 & Abs. \\
4.500 & \cellcolor{vFifty}2.94 & \cellcolor{vTwoHundred}\textbf{3.16} & \cellcolor{vFiveHundred}2.47 & \cellcolor{vThousand}1.70 & \cellcolor{vFifty}2.35 & \cellcolor{vTwoHundred}2.44 & \cellcolor{vFiveHundred}1.80 & \cellcolor{vThousand}1.07 & \cellcolor{vFifty}1.94 & \cellcolor{vTwoHundred}1.87 & \cellcolor{vFiveHundred}1.57 & \cellcolor{vThousand}1.34 & Abs. \\
5.500 & \cellcolor{vFifty}0.94 & \cellcolor{vTwoHundred}1.88 & \cellcolor{vFiveHundred}1.67 & \cellcolor{vThousand}0.71 & \cellcolor{vFifty}\textbf{3.30} & \cellcolor{vTwoHundred}\textbf{3.34} & \cellcolor{vFiveHundred}2.52 & \cellcolor{vThousand}2.22 & \cellcolor{vFifty}2.17 & \cellcolor{vTwoHundred}0.76 & \cellcolor{vFiveHundred}0.44 & \cellcolor{vThousand}1.74 & Abs. \\
5.524 & \cellcolor{vFifty}1.39 & \cellcolor{vTwoHundred}1.10 & \cellcolor{vFiveHundred}0.68 & \cellcolor{vThousand}0.00 & \cellcolor{vFifty}0.28 & \cellcolor{vTwoHundred}1.31 & \cellcolor{vFiveHundred}1.52 & \cellcolor{vThousand}1.64 & \cellcolor{vFifty}2.15 & \cellcolor{vTwoHundred}\textbf{3.15} & \cellcolor{vFiveHundred}\textbf{3.04} & \cellcolor{vThousand}2.16 & Em.\\
5.602 & \cellcolor{vFifty}2.49 & \cellcolor{vTwoHundred}2.29 & \cellcolor{vFiveHundred}1.67 & \cellcolor{vThousand}1.05 & \cellcolor{vFifty}0.77 & \cellcolor{vTwoHundred}0.56 & \cellcolor{vFiveHundred}0.55 & \cellcolor{vThousand}0.44 & \cellcolor{vFifty}\textbf{3.18} & \cellcolor{vTwoHundred}\textbf{3.20} & \cellcolor{vFiveHundred}2.08 & \cellcolor{vThousand}1.10 & Abs. \\
6.196 & \cellcolor{vFifty}1.50 & \cellcolor{vTwoHundred}1.48 & \cellcolor{vFiveHundred}1.62 & \cellcolor{vThousand}1.29 & \cellcolor{vFifty}\textbf{3.34} & \cellcolor{vTwoHundred}2.88 & \cellcolor{vFiveHundred}2.71 & \cellcolor{vThousand}2.64 & \cellcolor{vFifty}1.01 & \cellcolor{vTwoHundred}1.28 & \cellcolor{vFiveHundred}1.07 & \cellcolor{vThousand}1.39 & Em.\\
6.660 & \cellcolor{vFifty}1.35 & \cellcolor{vTwoHundred}2.66 & \cellcolor{vFiveHundred}\textbf{3.22} & \cellcolor{vThousand}\textbf{3.08} & \cellcolor{vFifty}0.88 & \cellcolor{vTwoHundred}1.63 & \cellcolor{vFiveHundred}2.56 & \cellcolor{vThousand}\textbf{3.01} & \cellcolor{vFifty}1.97 & \cellcolor{vTwoHundred}2.42 & \cellcolor{vFiveHundred}1.96 & \cellcolor{vThousand}1.06 & Em.\\
6.746 & \cellcolor{vFifty}2.75 & \cellcolor{vTwoHundred}\textbf{3.68} & \cellcolor{vFiveHundred}\textbf{3.77} & \cellcolor{vThousand}\textbf{3.12} & \cellcolor{vFifty}2.39 & \cellcolor{vTwoHundred}\textbf{3.21} & \cellcolor{vFiveHundred}\textbf{3.38} & \cellcolor{vThousand}2.97 & \cellcolor{vFifty}1.36 & \cellcolor{vTwoHundred}1.99 & \cellcolor{vFiveHundred}1.93 & \cellcolor{vThousand}1.29 & Em.\\
6.974 & \cellcolor{vFifty}1.97 & \cellcolor{vTwoHundred}2.58 & \cellcolor{vFiveHundred}\textbf{3.57} & \cellcolor{vThousand}2.56 & \cellcolor{vFifty}0.37 & \cellcolor{vTwoHundred}0.57 & \cellcolor{vFiveHundred}1.09 & \cellcolor{vThousand}0.33 & \cellcolor{vFifty}2.50 & \cellcolor{vTwoHundred}\textbf{3.40} & \cellcolor{vFiveHundred}\textbf{4.55} & \cellcolor{vThousand}\textbf{3.91} & Abs. \\
7.036 & \cellcolor{vFifty}1.15 & \cellcolor{vTwoHundred}2.42 & \cellcolor{vFiveHundred}\textbf{3.15} & \cellcolor{vThousand}\textbf{3.25} & \cellcolor{vFifty}0.66 & \cellcolor{vTwoHundred}2.28 & \cellcolor{vFiveHundred}\textbf{3.02} & \cellcolor{vThousand}2.99 & \cellcolor{vFifty}0.68 & \cellcolor{vTwoHundred}0.82 & \cellcolor{vFiveHundred}1.26 & \cellcolor{vThousand}1.40 & Em.\\
8.426 & \cellcolor{vFifty}2.49 & \cellcolor{vTwoHundred}\textbf{3.13} & \cellcolor{vFiveHundred}2.39 & \cellcolor{vThousand}2.26 & \cellcolor{vFifty}1.81 & \cellcolor{vTwoHundred}2.24 & \cellcolor{vFiveHundred}2.20 & \cellcolor{vThousand}2.23 & \cellcolor{vFifty}1.11 & \cellcolor{vTwoHundred}1.90 & \cellcolor{vFiveHundred}1.01 & \cellcolor{vThousand}0.89 & Abs. \\
8.538 & \cellcolor{vFifty}2.51 & \cellcolor{vTwoHundred}2.64 & \cellcolor{vFiveHundred}\textbf{3.09} & \cellcolor{vThousand}2.75 & \cellcolor{vFifty}1.20 & \cellcolor{vTwoHundred}1.68 & \cellcolor{vFiveHundred}2.02 & \cellcolor{vThousand}1.88 & \cellcolor{vFifty}1.52 & \cellcolor{vTwoHundred}1.72 & \cellcolor{vFiveHundred}2.32 & \cellcolor{vThousand}2.15 & Abs. \\
9.438 & \cellcolor{vFifty}0.80 & \cellcolor{vTwoHundred}1.70 & \cellcolor{vFiveHundred}1.16 & \cellcolor{vThousand}0.66 & \cellcolor{vFifty}0.32 & \cellcolor{vTwoHundred}0.00 & \cellcolor{vFiveHundred}0.47 & \cellcolor{vThousand}0.49 & \cellcolor{vFifty}\textbf{3.02} & \cellcolor{vTwoHundred}\textbf{3.63} & \cellcolor{vFiveHundred}\textbf{3.05} & \cellcolor{vThousand}2.04 & Abs. \\
9.494 & \cellcolor{vFifty}0.22 & \cellcolor{vTwoHundred}1.13 & \cellcolor{vFiveHundred}0.56 & \cellcolor{vThousand}0.00 & \cellcolor{vFifty}1.66 & \cellcolor{vTwoHundred}1.98 & \cellcolor{vFiveHundred}0.89 & \cellcolor{vThousand}0.42 & \cellcolor{vFifty}\textbf{3.61} & \cellcolor{vTwoHundred}1.59 & \cellcolor{vFiveHundred}0.41 & \cellcolor{vThousand}0.69 & Abs.\\\hline
Width (km/s) & \cellcolor{vFifty}50 & \cellcolor{vTwoHundred}200 & \cellcolor{vFiveHundred}500 & \cellcolor{vThousand}1000 & \cellcolor{vFifty}50 & \cellcolor{vTwoHundred}200 & \cellcolor{vFiveHundred}500 & \cellcolor{vThousand}1000 & \cellcolor{vFifty}50 & \cellcolor{vTwoHundred}200 & \cellcolor{vFiveHundred}500 & \cellcolor{vThousand}1000 & 
\enddata
\tablecomments{A row is included when at least one of the twelve fits has $\Delta C \geq 9$ (bold; approximately 3 local sigma). Each entry is $\sqrt{\Delta C}$ at the fixed energy in its row. The blue, green, yellow, and red columns indicate the 50, 200, 500, and 1000 km/s runs, respectively. Positive Gaussian normalizations are emission-like and negative normalizations are absorption. The type column summarizes only the above-threshold entries. Values are local and have not been corrected for the line-search trials.}
\end{deluxetable}


Table \ref{tab:three-line-search} shows all redshift values of our three-line search, together with the corresponding velocity and observed energy of the Fe XXV line, for which the local significance was above $1.5\sigma$. The bold entries represent $3\sigma$ detections. Only two values of the redshift (blueshift) produce significant detections. However, one of the two solutions is unphysical, as described in Section \ref{subsec:line_search}.

\begin{deluxetable}{c c c| c c c c| c c c c| c c c c}[h!]
\tablecaption{Three zgauss line search results\label{tab:three-line-search}}
\tabletypesize{\scriptsize}
\tablewidth{\textwidth}
\tablehead{
\colhead{$z$} & \colhead{$v/c$} & \colhead{Observed $E_{\mathrm{Fe\ XXV}}$} & \multicolumn4c{Time averaged} & \multicolumn4c{First 69 ks} & \multicolumn4c{Last 39 ks} \\
\colhead{} & \colhead{} & \colhead{(keV)} & \colhead{50} & \colhead{200} & \colhead{500} & \colhead{1000} & \colhead{50} & \colhead{200} & \colhead{500} & \colhead{1000} & \colhead{50} & \colhead{200} & \colhead{500} & \colhead{1000} \\
}
\startdata
+0.09872 & +0.0939 & 6.098 & \cellcolor{vFifty}0.66 & \cellcolor{vTwoHundred}0.00 & \cellcolor{vFiveHundred}0.00 & \cellcolor{vThousand}0.00 & \cellcolor{vFifty}0.00 & \cellcolor{vTwoHundred}0.00 & \cellcolor{vFiveHundred}0.00 & \cellcolor{vThousand}0.00 & \cellcolor{vFifty}1.68 & \cellcolor{vTwoHundred}0.85 & \cellcolor{vFiveHundred}0.52 & \cellcolor{vThousand}0.00 \\
+0.08731 & +0.0835 & 6.162 & \cellcolor{vFifty}0.93 & \cellcolor{vTwoHundred}0.29 & \cellcolor{vFiveHundred}0.00 & \cellcolor{vThousand}0.00 & \cellcolor{vFifty}1.56 & \cellcolor{vTwoHundred}0.20 & \cellcolor{vFiveHundred}0.00 & \cellcolor{vThousand}0.00 & \cellcolor{vFifty}0.00 & \cellcolor{vTwoHundred}0.00 & \cellcolor{vFiveHundred}0.14 & \cellcolor{vThousand}0.29 \\
+0.07925 & +0.0761 & 6.208 & \cellcolor{vFifty}0.00 & \cellcolor{vTwoHundred}0.00 & \cellcolor{vFiveHundred}0.00 & \cellcolor{vThousand}0.00 & \cellcolor{vFifty}1.72 & \cellcolor{vTwoHundred}0.00 & \cellcolor{vFiveHundred}0.00 & \cellcolor{vThousand}0.00 & \cellcolor{vFifty}0.00 & \cellcolor{vTwoHundred}0.00 & \cellcolor{vFiveHundred}0.00 & \cellcolor{vThousand}0.00 \\
+0.02698 & +0.0266 & 6.524 & \cellcolor{vFifty}1.63 & \cellcolor{vTwoHundred}0.00 & \cellcolor{vFiveHundred}0.00 & \cellcolor{vThousand}0.00 & \cellcolor{vFifty}0.00 & \cellcolor{vTwoHundred}0.00 & \cellcolor{vFiveHundred}0.00 & \cellcolor{vThousand}0.00 & \cellcolor{vFifty}0.58 & \cellcolor{vTwoHundred}0.12 & \cellcolor{vFiveHundred}0.00 & \cellcolor{vThousand}0.00 \\
+0.00359 & +0.0036 & 6.676 & \cellcolor{vFifty}0.00 & \cellcolor{vTwoHundred}0.00 & \cellcolor{vFiveHundred}0.00 & \cellcolor{vThousand}0.00 & \cellcolor{vFifty}0.00 & \cellcolor{vTwoHundred}0.00 & \cellcolor{vFiveHundred}0.00 & \cellcolor{vThousand}0.00 & \cellcolor{vFifty}1.50 & \cellcolor{vTwoHundred}1.05 & \cellcolor{vFiveHundred}0.09 & \cellcolor{vThousand}0.71 \\
-0.00089 & -0.0009 & 6.706 & \cellcolor{vFifty}1.67 & \cellcolor{vTwoHundred}2.94 & \cellcolor{vFiveHundred}2.81 & \cellcolor{vThousand}1.62 & \cellcolor{vFifty}0.00 & \cellcolor{vTwoHundred}0.21 & \cellcolor{vFiveHundred}0.00 & \cellcolor{vThousand}0.00 & \cellcolor{vFifty}2.05 & \cellcolor{vTwoHundred}\textbf{3.46} & \cellcolor{vFiveHundred}\textbf{3.96} & \cellcolor{vThousand}\textbf{3.24} \\
-0.03929 & -0.0401 & 6.974 & \cellcolor{vFifty}1.07 & \cellcolor{vTwoHundred}1.81 & \cellcolor{vFiveHundred}2.93 & \cellcolor{vThousand}1.77 & \cellcolor{vFifty}0.00 & \cellcolor{vTwoHundred}0.00 & \cellcolor{vFiveHundred}0.00 & \cellcolor{vThousand}0.00 & \cellcolor{vFifty}1.70 & \cellcolor{vTwoHundred}2.74 & \cellcolor{vFiveHundred}\textbf{3.99} & \cellcolor{vThousand}\textbf{3.30} \\
-0.04477 & -0.0458 & 7.014 & \cellcolor{vFifty}0.00 & \cellcolor{vTwoHundred}0.00 & \cellcolor{vFiveHundred}0.00 & \cellcolor{vThousand}0.00 & \cellcolor{vFifty}0.00 & \cellcolor{vTwoHundred}0.00 & \cellcolor{vFiveHundred}0.00 & \cellcolor{vThousand}0.00 & \cellcolor{vFifty}1.62 & \cellcolor{vTwoHundred}0.81 & \cellcolor{vFiveHundred}0.00 & \cellcolor{vThousand}0.00 \\
-0.05018 & -0.0514 & 7.054 & \cellcolor{vFifty}0.30 & \cellcolor{vTwoHundred}0.47 & \cellcolor{vFiveHundred}0.30 & \cellcolor{vThousand}0.08 & \cellcolor{vFifty}1.65 & \cellcolor{vTwoHundred}1.24 & \cellcolor{vFiveHundred}0.45 & \cellcolor{vThousand}0.26 & \cellcolor{vFifty}0.00 & \cellcolor{vTwoHundred}0.00 & \cellcolor{vFiveHundred}0.00 & \cellcolor{vThousand}0.00 \\
-0.07099 & -0.0735 & 7.212 & \cellcolor{vFifty}0.79 & \cellcolor{vTwoHundred}0.00 & \cellcolor{vFiveHundred}0.00 & \cellcolor{vThousand}0.00 & \cellcolor{vFifty}1.53 & \cellcolor{vTwoHundred}0.83 & \cellcolor{vFiveHundred}0.37 & \cellcolor{vThousand}0.00 & \cellcolor{vFifty}1.66 & \cellcolor{vTwoHundred}0.48 & \cellcolor{vFiveHundred}0.00 & \cellcolor{vThousand}0.00 \\
-0.08744 & -0.0912 & 7.342 & \cellcolor{vFifty}0.79 & \cellcolor{vTwoHundred}0.83 & \cellcolor{vFiveHundred}0.68 & \cellcolor{vThousand}0.23 & \cellcolor{vFifty}1.66 & \cellcolor{vTwoHundred}1.08 & \cellcolor{vFiveHundred}0.63 & \cellcolor{vThousand}0.60 & \cellcolor{vFifty}0.00 & \cellcolor{vTwoHundred}0.00 & \cellcolor{vFiveHundred}0.00 & \cellcolor{vThousand}0.00 \\
-0.09288 & -0.0972 & 7.386 & \cellcolor{vFifty}1.41 & \cellcolor{vTwoHundred}0.00 & \cellcolor{vFiveHundred}0.00 & \cellcolor{vThousand}0.00 & \cellcolor{vFifty}1.74 & \cellcolor{vTwoHundred}0.17 & \cellcolor{vFiveHundred}0.00 & \cellcolor{vThousand}0.00 & \cellcolor{vFifty}0.00 & \cellcolor{vTwoHundred}0.00 & \cellcolor{vFiveHundred}0.00 & \cellcolor{vThousand}0.00 \\
-0.12189 & -0.1293 & 7.630 & \cellcolor{vFifty}0.59 & \cellcolor{vTwoHundred}0.61 & \cellcolor{vFiveHundred}0.22 & \cellcolor{vThousand}0.39 & \cellcolor{vFifty}1.91 & \cellcolor{vTwoHundred}1.39 & \cellcolor{vFiveHundred}0.93 & \cellcolor{vThousand}0.93 & \cellcolor{vFifty}0.00 & \cellcolor{vTwoHundred}0.00 & \cellcolor{vFiveHundred}0.00 & \cellcolor{vThousand}0.00 \\
-0.12487 & -0.1326 & 7.656 & \cellcolor{vFifty}1.58 & \cellcolor{vTwoHundred}0.00 & \cellcolor{vFiveHundred}0.48 & \cellcolor{vThousand}0.83 & \cellcolor{vFifty}1.83 & \cellcolor{vTwoHundred}1.34 & \cellcolor{vFiveHundred}1.27 & \cellcolor{vThousand}1.29 & \cellcolor{vFifty}0.00 & \cellcolor{vTwoHundred}0.00 & \cellcolor{vFiveHundred}0.00 & \cellcolor{vThousand}0.00 \\
-0.12829 & -0.1364 & 7.686 & \cellcolor{vFifty}0.68 & \cellcolor{vTwoHundred}0.00 & \cellcolor{vFiveHundred}0.36 & \cellcolor{vThousand}0.66 & \cellcolor{vFifty}1.56 & \cellcolor{vTwoHundred}0.00 & \cellcolor{vFiveHundred}0.00 & \cellcolor{vThousand}0.28 & \cellcolor{vFifty}0.00 & \cellcolor{vTwoHundred}0.32 & \cellcolor{vFiveHundred}0.60 & \cellcolor{vThousand}0.02 \\
-0.13793 & -0.1473 & 7.772 & \cellcolor{vFifty}0.00 & \cellcolor{vTwoHundred}0.00 & \cellcolor{vFiveHundred}0.00 & \cellcolor{vThousand}0.00 & \cellcolor{vFifty}0.00 & \cellcolor{vTwoHundred}0.00 & \cellcolor{vFiveHundred}0.00 & \cellcolor{vThousand}0.00 & \cellcolor{vFifty}1.82 & \cellcolor{vTwoHundred}0.40 & \cellcolor{vFiveHundred}0.00 & \cellcolor{vThousand}0.35 \\
-0.15383 & -0.1655 & 7.918 & \cellcolor{vFifty}1.16 & \cellcolor{vTwoHundred}1.06 & \cellcolor{vFiveHundred}0.00 & \cellcolor{vThousand}0.00 & \cellcolor{vFifty}1.16 & \cellcolor{vTwoHundred}0.43 & \cellcolor{vFiveHundred}0.00 & \cellcolor{vThousand}0.00 & \cellcolor{vFifty}1.65 & \cellcolor{vTwoHundred}1.12 & \cellcolor{vFiveHundred}0.00 & \cellcolor{vThousand}0.00 \\
-0.15914 & -0.1716 & 7.968 & \cellcolor{vFifty}0.64 & \cellcolor{vTwoHundred}0.00 & \cellcolor{vFiveHundred}0.37 & \cellcolor{vThousand}0.83 & \cellcolor{vFifty}0.03 & \cellcolor{vTwoHundred}0.00 & \cellcolor{vFiveHundred}0.69 & \cellcolor{vThousand}1.11 & \cellcolor{vFifty}1.71 & \cellcolor{vTwoHundred}1.07 & \cellcolor{vFiveHundred}0.56 & \cellcolor{vThousand}0.00 \\
\enddata
\tablecomments{A row is included when at least one of the twelve three-zgauss fits has a local significance of at least 1.5 sigma ($\Delta C \geq 5.41$); bold entries reach 3 local sigma ($\Delta C \geq 13.22$). The first two columns give the shared zgauss redshift and its special-relativistic line-of-sight velocity, $v/c=[(1+z)^2-1]/[(1+z)^2+1]$; negative values indicate blueshift. The 6.952- and 6.973-keV components share this redshift and width, while their normalizations are tied 2:1. Each entry is the one-sided Gaussian-equivalent local significance obtained from a conservative $\chi^2$ distribution with two degrees of freedom, for the two free line normalizations at fixed redshift. All line normalizations were constrained to be non-positive, so all candidates are absorption. Values are local and have not been corrected for the line-search trials.}
\end{deluxetable}

\section{Spectra} \label{sec:spectra_plots}

\begin{sidewaysfigure}
    \centering
    \includegraphics[width=0.95\linewidth]{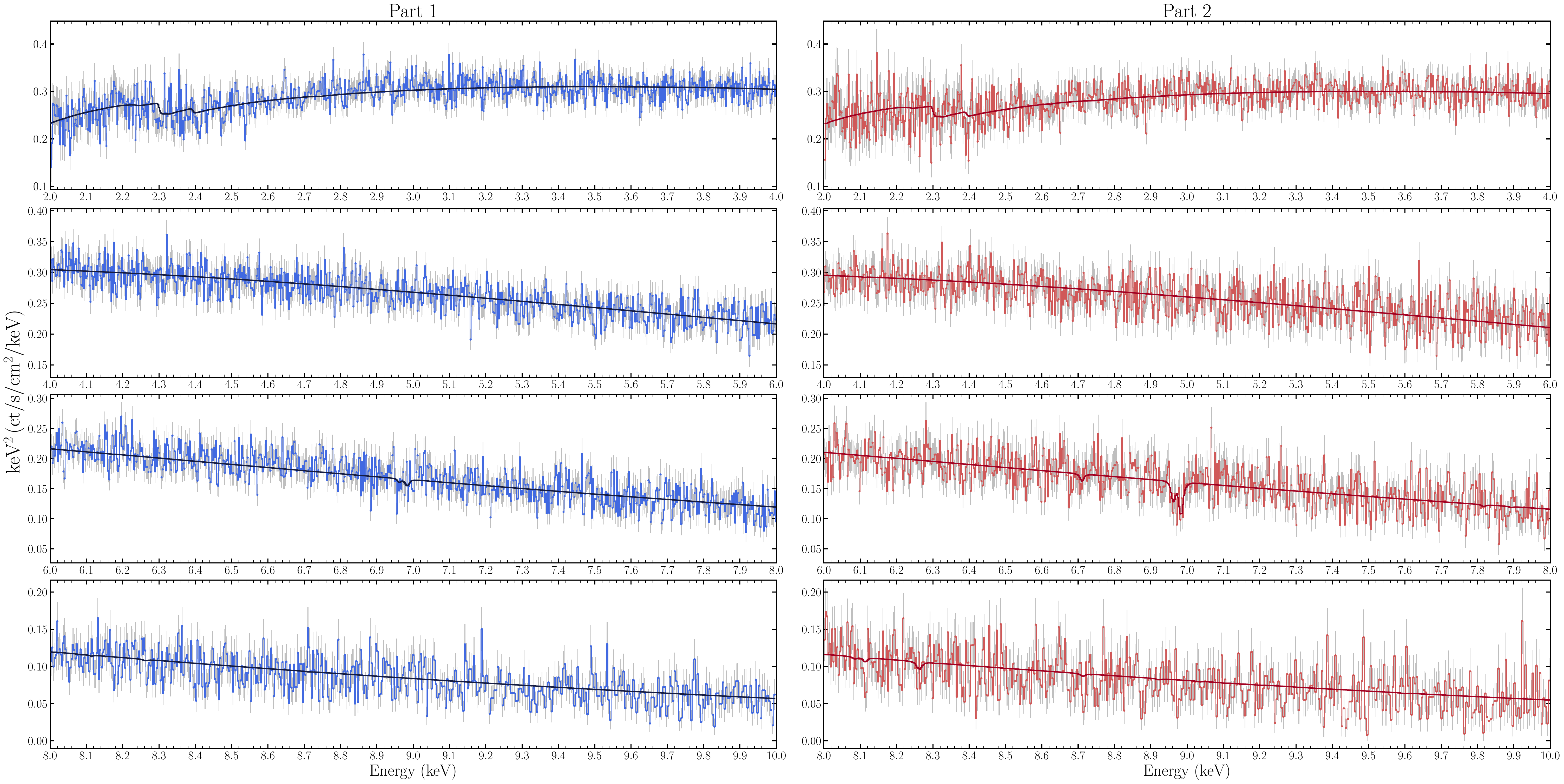}
    \caption{Part 1 (blue, left) and Part 2 (red, right) Resolve spectra of 4U 1957+115 in the 2-4, 4-6, 6-8, and 8-10 keV, respectively. The solid lines indicate the respective best-fit models for the two spectra. Other than the Fe XXVI line around 7 keV, no narrow absorption or emission features stand out.}
    \label{fig:full_spec_parts1and2}
\end{sidewaysfigure}

Figure \ref{fig:full_spec_parts1and2} shows the full Part 1 and Part 2 spectra, in the entire 2-10 keV Resolve band. Other than the Fe XXVI line in the Part 2 spectrum, no other narrow emission or absorption features are immediately noticeable. This is corroborated by the result of the line-search algorithm (Section \ref{subsec:line_search} and Appendix \ref{sec:line_search_results}).

\begin{figure}
    \centering
    \includegraphics[width=0.75\linewidth]{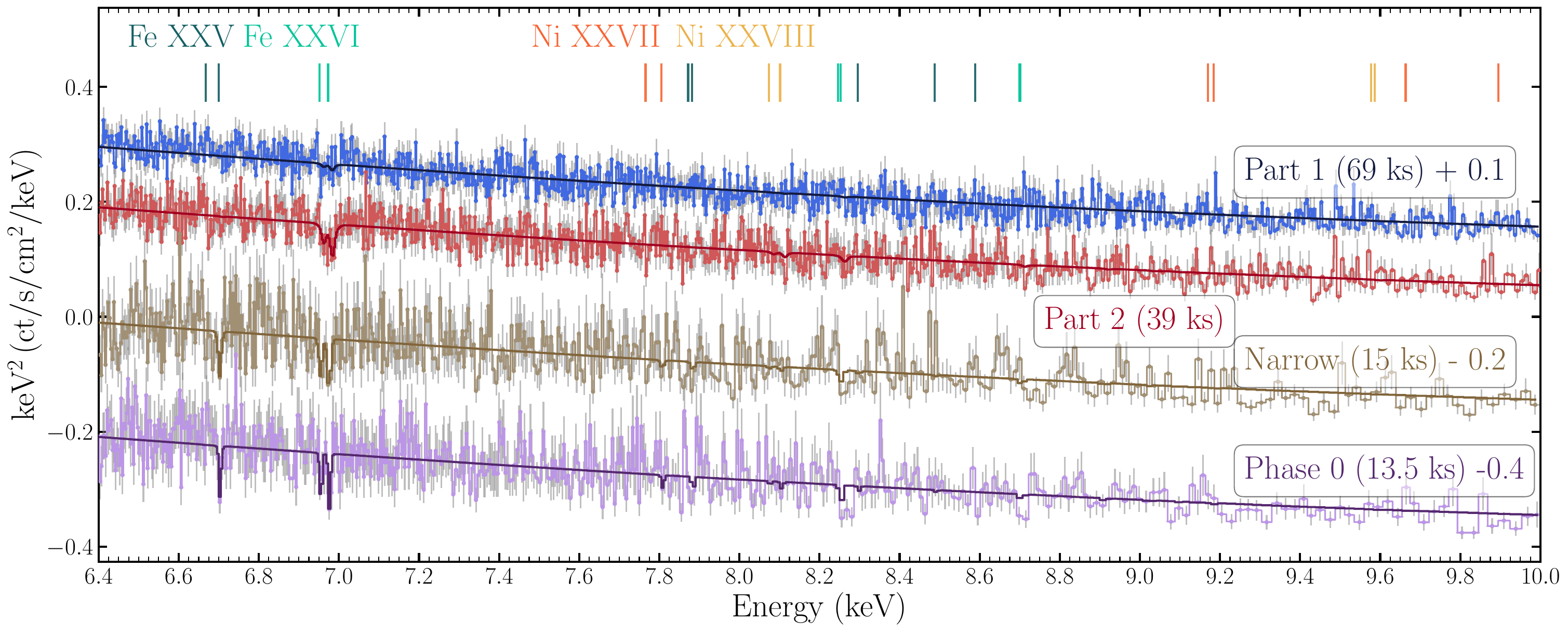}
    \caption{6.4-10 keV Resolve spectra, generated as splitting the observation in the same way as the one described in Figure \ref{fig:Fe_band_resolved}. The color scheme follows that in Figure \ref{fig:Fe_band_resolved}. The colored lines at the top show the expected positions of strong transitions of Fe XXV, Fe XXVI, Ni XXVII, and Ni XXVIII. The spectra were rebinned for visual clarity.}
    \label{fig:time_split_wide}
\end{figure}

Figure \ref{fig:time_split_wide} shows the spectra presented in Figure \ref{fig:Fe_band_resolved}, but in the 6.4-10 keV band, using the same coloring and naming conventions.

\section{Corner plots} \label{sec:corner_plots}

\begin{figure}
    \centering
    \includegraphics[width=0.8\linewidth]{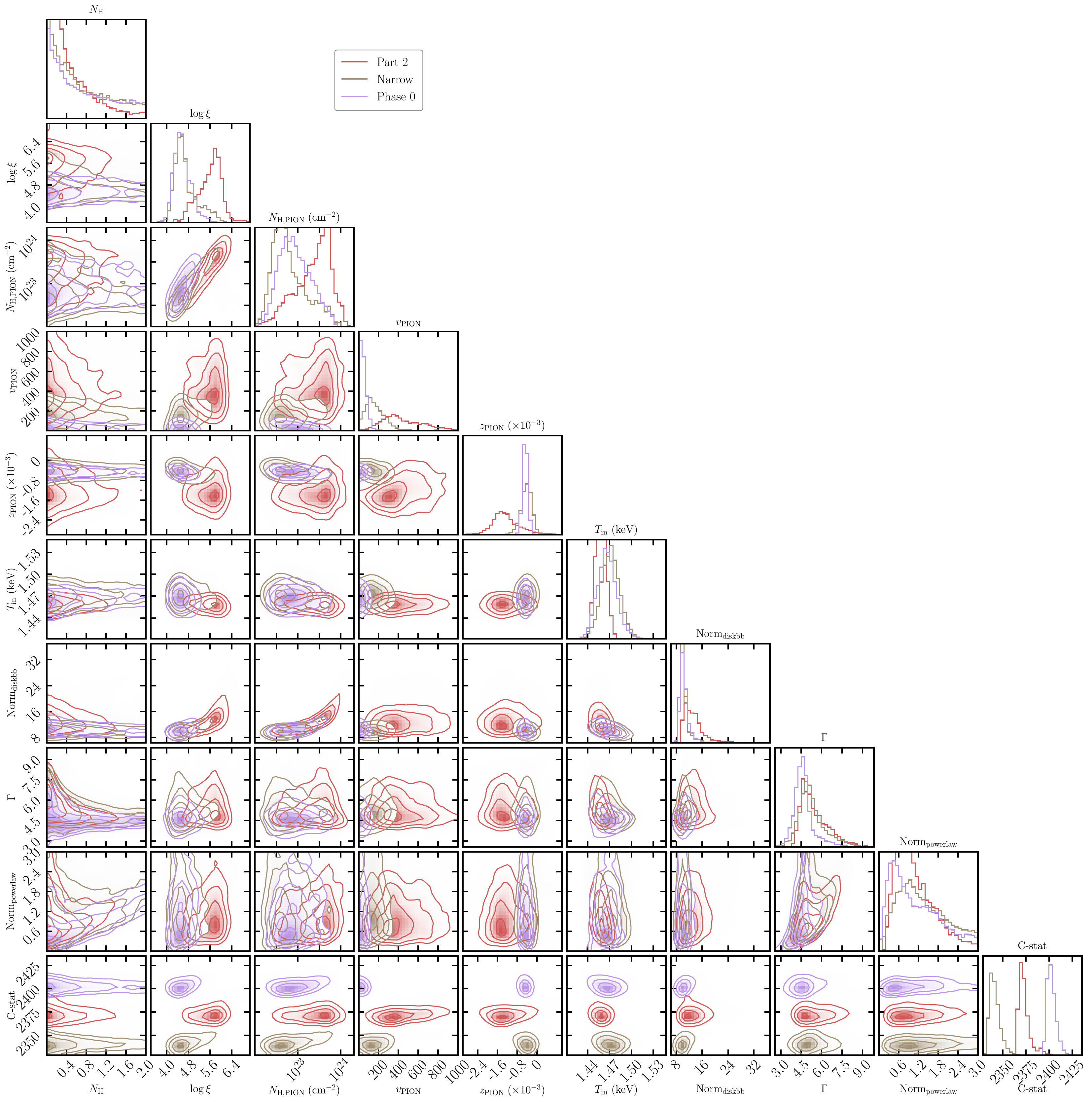}
    \caption{Corner plot resulting from the MCMC analysis of the Part 2, Narrow, and Phase 0 spectra. The color convention was maintained as in Figures \ref{fig:Fe_band_resolved} and \ref{fig:time_split_wide}.}
    \label{fig:corner}
\end{figure}

Figure \ref{fig:corner} shows the corner plot resulting from the MCMC analysis of the Part 2, Narrow, and Phase 0 spectra. All the parameters in the \texttt{TBabs $\times$ pion $\times$ (diskbb+powerlaw)} model are shown. A strong correlation is noticeable between the ionization $\log\xi$ and the column density of the \texttt{pion} component $N_{\rm H, PION}$.

\end{document}